\documentclass[a4paper,11pt]{article}

\usepackage[T1]{fontenc}
\usepackage{comment}
\usepackage{fixltx2e}
\usepackage[table,xcdraw]{xcolor}
\usepackage{jcappub}
\usepackage{float}
\usepackage[scr=boondox]{mathalfa}
\usepackage{bm}
\usepackage{stmaryrd}
\usepackage{bbold}
\usepackage{bbm}
\usepackage[normalem]{ulem}
\usepackage{amsmath}
\usepackage{geometry}
\usepackage{cancel}
\usepackage{graphbox}
\usepackage{subfig}
\usepackage{booktabs}
\usepackage{etoolbox} % needed for \perm
\usepackage{array}
\newcommand{\be}{\begin{equation}}
\newcommand{\ee}{\end{equation}}

\newcommand{\bea}{\begin{eqnarray}}
\newcommand{\eea}{\end{eqnarray}}

\def\ba#1\ea{\begin{align}#1\end{align}}
\def\bg#1\eg{\begin{gather}#1\end{gather}}

\newcommand{\refeq}[1]{Eq.~(\ref{eq:#1})}

\newcommand{\reffig}[1]{Fig.~\ref{fig:#1}}

\newcommand{\reftab}[1]{Tab.~\ref{tab:#1}}
\newcommand{\refsec}[1]{Sec.~\ref{sec:#1}}
\newcommand{\refapp}[1]{App.~\ref{app:#1}}

\newcommand{\Z}{\mathcal{Z}}

\newcommand{\Del}{\mathcal{D}}
\newcommand{\Prob}{\mathcal{P}}

\renewcommand{\P}{\mathcal{P}}

\newcommand{\unit}{{\mathbb{1}}}

\DeclareMathOperator{\tr}{tr}

\renewcommand{\v}[1]{\bm{#1}}

\newcommand{\vx}{\v{x}}

\newcommand{\vk}{\v{k}}
\newcommand{\vq}{\v{q}}
\newcommand{\vp}{\v{p}}

\newcommand{\vs}{\nonumber\\}

\renewcommand{\L}{\Lambda}

\newcommand{\kmax}{k_\text{max}}

\newcommand{\dg}{\delta_g}
\newcommand{\din}{\delta_{\rm in}}
\newcommand{\dlin}{\delta^{(1)}}

\newcommand{\dshell}{\delta_{\rm shell}^{(1)}}

\newcommand{\dgdet}{\delta_{g,\rm det}}

\newcommand{\dinL}{\delta_{\rm in,\Lambda}}
\newcommand{\dlinL}{\delta^{(1)}_\Lambda}
\newcommand{\dlinLin}{\delta^{(1)}_{\Lambda_\mathrm{in}}}

\newcommand{\hdg}{\hat{\delta}_g}

\newcommand{\hdgkmax}{\hat{\delta}_{g,\kmax}}

\newcommand{\tO}[1]{\widetilde{#1}}
\newcommand{\tObot}[1]{\widetilde{#1}^\bot}

\newcommand{\dirac}{\delta_{\rm D}}

\newcommand{\diracc}{\delta_{\rm D}}
\newcommand{\diracpi}{\hat{\delta}_{\rm D}}

\newcommand{\Plin}{P_{\rm L}}

\newcommand{\Ds}{\Delta\sigma^2}

\newcommand{\Zcorr}[1]{\zeta^{(#1)}}

\newcommand{\lapl}{\nabla^2}

\newcommand{\npt}{n \text{-pt}}

\newcommand{\perm}[1]{%
  \expandafter\ifstrempty\expandafter{#1}
  {\mbox{perm.}}
  {\mbox{$#1$ perm.}}%
}

\makeatletter
\newlength{\apb@width}
\newcommand{\autoparbox}[2][c]{%
  \settowidth{\apb@width}{#2}%
  \parbox[#1]{\apb@width}{#2}%
}

\makeatother

\title{Connecting field-level and n-point function measurements of galaxy bias}

\author[a,d]{Hsiang-Ming Huang,}
\emailAdd{hmhuang@mpa-garching.mpg.de}
\author[a,b,c]{Henrique Rubira,}
\author[d]{Fabian Schmidt}
\affiliation[a]{University Observatory, Faculty of Physics, Ludwig-Maximilians-Universit\"at, Scheinerstr. 1, D-81679 München, Germany}
\affiliation[b]{Kavli Institute for Cosmology Cambridge, Madingley Road, Cambridge CB3 0HA, UK}
\affiliation[c]{Centre for Theoretical Cosmology, Department of Applied Mathematics and Theoretical Physics
	University of Cambridge, Wilberforce Road, Cambridge, CB3 0WA, UK}
    
\affiliation[d]{Max-Planck-Institut f\"{u}r Astrophysik, Karl-Schwarzschild-Str. 1, 85748 Garching, Germany}

\abstract{
  Precise measurements of galaxy bias parameters are essential for extracting cosmological information from galaxy surveys.
  Field-level analyses infer bias coefficients at a finite cutoff $\Lambda$, typically related to the lattice resolution, while $n$-point function analyses typically adopt renormalization conditions defined in the low-$k$ limit. In this work, we use the renormalization group (RG) to relate field-level measurements at different cutoffs and connect them to $n$-point function measurements. We generate mock galaxy fields with a Lagrangian perturbation theory forward model and a galaxy bias expansion, and then measure the bias parameters across a range of cutoffs. We show that the RG predictions describe the measured running and, in the limit $\Lambda\to 0$, agree with the (tree-level) $n$-point function measurements of first-order $b_\delta$ and second-order $b_{\delta^2}, b_{\mathcal{G}_2}$. We also measure the running on halo catalogs based on full $N$-body simulations and show that the RG prediction describes the measurements well within the predicted systematic errors. Our results establish the RG flow as a natural formulation for relating galaxy bias measurements at different cutoffs and connecting the schemes used in field-level and $n$-point function analyses.
} 


\begin{document}

\maketitle
\flushbottom

%%%%%%%%%%%%%%%%%%%%%%%%%%%%%%%%%%%%%%%%%%%%%%%%%%%%%%%%%%%%%%%%%%%%%%%%%%%
%%%%%%%%%%%%%%%%%%%%%%%%%%%%%%%%%%%%%%%%%%%%%%%%%%%%%%%%%%%%%%%%%%%%%%%%%%%

\section{Introduction}\label{sec:introduction}

Increasingly precise surveys of large-scale structure (LSS) offer new opportunities to constrain dark matter, dark energy, gravity, and the initial conditions of the Universe.
Extracting more information requires a consistent theoretical description that connects observable tracers, such as galaxies, to the matter distribution predicted by cosmological models.
Within the effective field theory (EFT) and galaxy bias expansion frameworks (see \cite{desjacques_large-scale_2018} for a review), the deterministic component of the galaxy overdensity field can be perturbatively written as 
\begin{equation}\label{eq:bias_expansion}
    \delta_{g,{\rm det}}(\vx,\tau) =\sum_O b_O(\tau) O(\vx,\tau),
\end{equation}
where the bias coefficients $b_O$ and operators $O$ capture the most general expression allowed by gravity, with the operators $O(\vx,\tau)=O[\delta^{(1)}(\vx,\tau)]$ being functionals of the linear matter density field $\delta^{(1)}(\vx,\tau)$.
Describing stochastic fluctuations around the deterministic \refeq{bias_expansion} requires an additional random field $\epsilon$ with separate effective coupling constants \cite{rubira2025nongaussiangalaxystochasticitynoisefield}.

Tracer $n$-point functions (e.g., the prediction for the galaxy power spectrum) are then computed from \refeq{bias_expansion}, involving correlators of operators. These in general contain ultraviolet (UV)-sensitive loop contributions (integrals) which need to be regularized, and \textit{renormalization conditions} need to be introduced to provide well-defined predictions. The most commonly chosen conditions on a given operator $O$ are \cite{assassi_renormalized_2014}
\be\label{eq:renormalization_conditions}
\lim_{q_i\to 0}\langle [O]_{\vq} \dlin_{\vq_1}\cdots\dlin_{\vq_n}\rangle= \langle O_{\vq} \dlin_{\vq_1}\cdots\dlin_{\vq_n}\rangle_{\rm tree},
\ee
where $O$ denotes the bare operator, and $[O]$ is its renormalized counterpart. Since these conditions are defined via tree-level $n$-point functions on the right-hand side, we refer to the $[O]$ as \emph{$\npt$-renormalized operators}, and the corresponding bias coefficients as \emph{$\npt$ bias parameters}. Assuming that the complete set of bias operators is employed, the counterterms involved in the construction of $[O]$ are the bias operators themselves, both at leading and higher order in derivatives.

More recently, Ref.~\cite{Rubira:2023vzw} introduced an alternative, finite ``lattice''-cutoff approach based on the renormalization group (RG) \cite{Wilson:1971bg,POLCHINSKI1984269}. In this framework, ordinary differential equations describe how the bias parameters evolve as the cutoff changes. Both the lattice cutoff and the renormalized bias framework based on \refeq{renormalization_conditions} are expected to match when the lattice cutoff is sent to zero, corresponding to just matching tree-level correlators \cite{Rubira:2023vzw}.
The lattice-cutoff RG framework has since been extended to include stochastic terms \cite{rubira_renormalization_2024}, primordial non-Gaussianity \cite{nikolis_renormalization_2024}, two-loop contributions \cite{Bakx:2025cvu}, and higher-derivative operators \cite{bakx2026}.

A fundamental alternative approach to measuring bias using $n$-point correlation functions is to employ the measured and predicted tracer fields, $\delta_g$ and $\delta_{g,\rm det}$, respectively, directly \cite{Abidi_2018,lazeyras_beyond_2018,Schmittfull:2018yuk,lazeyras_assembly_2021,barreira_galaxy_2021,ivanov2024fullshapeanalysissimulationbasedpriors,Obuljen_2023}. In case of simulated tracers, one can employ the initial conditions used in the simulations for the prediction, thus canceling cosmic variance to the largest degree possible. However, {\it any} field-level measurement introduces a numerical cutoff scale, making bias parameters inferred at different cutoffs or in different studies difficult to compare. One of the advantages of the lattice cutoff scheme introduced by \cite{Rubira:2023vzw} is that it naturally connects to perturbative field-level bias measurements.

In this paper, we establish in detail the connection between field-level bias measurements and the finite-cutoff RG predictions, measure for the first time the running of bias parameters at the field level, and compare it with the RG predictions of \cite{Rubira:2023vzw}. 
We use \texttt{LEFTfield} \cite{Schmidt_2021}, a Lagrangian EFT forward model that evolves initial conditions using Lagrangian perturbation theory (LPT) and constructs the tracer density field using the galaxy bias expansion.
We also examine how bias parameters defined at a finite cutoff relate to those obtained from renormalized $n$-point function measurements \cite{assassi_renormalized_2014} in the limit of zero cutoff. 
We show that the renormalization group equations provide a natural way to relate bias parameters at different reference scales, using them to predict the bias values at other cutoffs.

The paper is organized as follows. In \refsec{theory}, we review the RG framework for galaxy bias and discuss the theoretical uncertainties associated with its truncation.
In \refsec{simulation_measurement}, we describe the simulation setup, including the grids and filters, and both the field-level likelihood derived from the partition function and the $n$-point function measurement of the bias parameters.
We present the results and relation to the transfer-function approach proposed in \cite{Schmittfull:2018yuk} in \refsec{results}. We conclude in \refsec{conclusion}.

\subsection*{Dictionary and notation}

We introduce the following notation used throughout the paper:
\begin{itemize}
    \item $\int_{\vp_1,...,\vp_n}=\int\frac{d^3 p_1}{(2\pi)^3}...\int\frac{d^3 p_n}{(2\pi)^3}$ for momentum integrals.
    \item $\diracpi(\vk)=(2\pi)^3\diracc (\vk)$ for a reduced 3D Dirac delta distribution.
    \item $f_\Lambda(\vk)=W_\Lambda(k)f(\vk)$ for a sharp-$k$ filter at the scale $\Lambda$ applied to a field $f$ in momentum space.
    \item $\langle O(\vk_1)...O(\vk_n)\rangle=\diracpi \langle O(\vk_1)...O(\vk_n)\rangle'$ for the shorthand prime notation for correlators.
\end{itemize}

\paragraph{Scales}
\begin{itemize}
    \item $\Lambda$ denotes the smoothing scale in the initial conditions.
    \item $\Lambda_*$:is the \textit{fixed} reference value of the smoothing scale, taken at $\Lambda_* = 0.1 \, h/{\rm Mpc}$ unless stated otherwise.
    \item $k_\mathrm{max}$: is the highest wavenumber considered in an analysis, and we always assume $k_\mathrm{max}\leq\Lambda$.
\end{itemize}
Both $\L$ and $k_{\rm max}$ are always chosen smaller than the non-linear scale where perturbation theory breaks down. Note that while $\L$ can be chosen to be non-perturbative in the numerical inference, a controlled prediction for the running requires that $\L$ is in the perturbative regime.

\paragraph{Operators}
\begin{itemize}
    \item $O[\delta^{(1)}_\Lambda]$: operators constructed from the linear matter density $\dlinL$ smoothed at a scale $\Lambda$. For an operator with the leading perturbative order ${\rm Ord}(O)$ (often denoted as $n(O)$), it is constructed with kernels $K_O^{(n)}$ as follows
    \be\label{eq:operator_kernel}
    O[\dlinL](\vk)=\sum_{n={\rm Ord}(O)}^\infty\int_{\vp_1,...,\vp_n}\diracpi (\vk-\vp_{1...n})K^{(n)}_O(\vp_1,...,\vp_n)\dlinL(\vp_1)...\dlinL(\vp_n).
    \ee 
    Note that the kernels $K^{(n)}_O$ are the combination of matter kernels and galaxy bias kernels.   We also use $b_{O^{[n]}}$ to refer to bias coefficients that start contributing at $n={\rm Ord}(O)$. We adopt the following operator basis, up to third-order: $\{\delta,\delta^2,\mathcal{G}_2,\delta^3,\delta \mathcal{G}_2,\mathcal{G}_3,\Gamma_3, \nabla^2 \delta\}$; see \cite{assassi_renormalized_2014} for the definitions.
    \item $[O]$: the $n$-point function renormalized operator $O$, defined through the renormalization conditions of \refeq{renormalization_conditions}.
\end{itemize}

\paragraph{Initial conditions and time evolution}\mbox{}\\

We consider Gaussian adiabatic initial conditions $\delta_\mathrm{in}(\vx)$ following a distribution
\be\label{eq:gaussian_ic_pdf}
\Prob[\din|\{\theta\}]=\left[ \prod_{\vk}^\Lambda 2\pi P_\mathrm{in}(k|\theta) \right]^{-1/2}\exp\left[ -\frac{1}{2}\int_{\vk}^\Lambda\frac{|\din(\vk)|^2}{P_\mathrm{in}(k|\theta)}\right],
\ee
where $\{\theta\}$ are cosmological parameters. 
We also assume that the late-time matter field is described by a single scale-independent growing mode 
\be 
\dlin(\vx,\tau)=D(\tau)\din(\vx).
\ee 
The initial and linearly evolved power spectrum are therefore related by 
\begin{equation}\label{eq:pk_lin}
    \Plin(k,\tau)=\langle \delta^{(1)}(\vk,\tau)\delta^{(1)}(\vk',\tau)\rangle'=D^2(\tau)P_\mathrm{in}(k),
\end{equation}
and the time dependence can be absorbed into the normalization of the linear field. We further adopt the EdS approximation \cite{bernardeau_large-scale_2002,Garny:2020ilv} for the higher-order perturbative fields' evolution. This does not affect the results noticeably. 
The variance of the linear density field at a scale $\Lambda$ is given by
\be 
\sigma_\Lambda^2(\tau)\equiv\int^\Lambda_{\vp} \Plin(p,\tau) \,.
\ee
We further define $\Ds$ as 
\be\label{eq:Ds}
\Ds(\tau) \equiv \sigma_{\Lambda_*}^2-\sigma^2_\Lambda = \int_{\Lambda}^{\Lambda_*} \Plin(p,\tau).
\ee 
In the following, we will generally drop the explicit time argument. It is always set to the observation time of the tracers considered, which we state when defining the datasets employed.

%%%%%%%%%%%%%%%%%%%%%%%%%%%%%%%%%%%%%%%%%%%%%%%%%%%%%%%%%%%%%%%%%%%%%%%%%%%
%%%%%%%%%%%%%%%%%%%%%%%%%%%%%%%%%%%%%%%%%%%%%%%%%%%%%%%%%%%%%%%%%%%%%%%%%%%
\section{The renormalization group (RG) for LSS}\label{sec:theory}

In this section, we derive the renormalization group equations that give the bias parameters' dependence on the cutoff-scale adopting the partition-function viewpoint for the derivation. We also discuss the truncation of these equations up to a fixed perturbative order.

\subsection{The partition function for the galaxy density}

The partition function for the galaxy density field can be written as \cite{cabass_likelihood_2020,Rubira:2023vzw,rubira_renormalization_2024,rubira2025nongaussiangalaxystochasticitynoisefield} 
\begin{equation}\label{eq:partition_function}
    \mathcal{Z}[J_{\kmax}]=\int\mathcal{D}\dlinL\, \mathcal{P}[\dlinL]\exp\left(S_\mathrm{eff}[\dlinL,J_{\kmax}]\right),
\end{equation}
with the effective action
\be\label{eq:effective_action}
    S_\mathrm{eff}[\dlinL,J_{\kmax}]\equiv\sum_{m=1}^\infty\sum_{\unit,O}\frac{C^{\{m\}}_O(\Lambda)}{m!}\int_{\vx}[J_{\kmax}(\vx)]^m\,O[\dlinL](\vx),
\ee
which sums local-in-space products of the current $J_{\kmax}$, which by construction has support only up to $\kmax < \L$, and operators $O$. We use $J=J_{\kmax}$ for clarity of notation in the following. The deterministic bias parameters $b_O = C^{\{1\}}_O$ appearing in \refeq{bias_expansion}, are given by $J^1$ terms, the amplitude of Gaussian stochastic contributions is given by $P_{\epsilon,O} = C^{\{2\}}_O$, and $C^{\{m\geq 3\}}_O$ encodes non-Gaussian stochastic contributions, e.g. $B_{\epsilon,\unit}=C_\unit^{\{3\}}$. The partition function is closed under the RG flow, in the sense that a shift $\Lambda \to \Lambda + \lambda$ is absorbed into the coefficients $C^{\{m\}}_O(\Lambda)$ \cite{rubira_renormalization_2024}. We neglect higher-derivative-in-current contributions such as $\lapl J$ or $\partial_i J\,\partial^i J$ (see \cite{bakx2026}). For stochastic contributions ($m>2$), it is also necessary to include a constant operator $\unit$ in the basis in \refeq{effective_action}, which gives a generalized shot-noise contribution to $n$-point functions. 

The $n$-point functions are determined through functional derivatives of the partition function with respect to the current,
\be
\langle \dg(\vk_1)\cdots\dg(\vk_n)\rangle
=
\frac{1}{\mathcal{Z}[0]}
\left.
\frac{\delta^n\mathcal{Z}[J]}
{\delta J(-\vk_1)\cdots\delta J(-\vk_n)}
\right|_{J=0}.
\ee
Explicitly, the first three $n$-point functions read \cite{rubira_renormalization_2024}
\ba
\langle \dg(\vk) \rangle &= b_{\unit} \diracpi(\vk) = 0\,,  \nonumber\\
\langle \dg(\vk_1) \dg(\vk_2) \rangle &= b_{O_a} b_{O_b} \langle O_a(\vk_1) O_b(\vk_2) \rangle + P_{\epsilon,\unit}\, \diracpi(\vk_{12})\,, \label{eq:dd_general}\\
\langle \dg(\vk_1) \dg(\vk_2) \dg(\vk_3) \rangle &= b_{O_{a}} b_{O_{b}} b_{O_c} \langle O_a(\vk_1) O_b(\vk_2)O_c(\vk_3) \rangle \nonumber\\
&\quad+ \left( b_{O_a} P_{\epsilon,O_{b}} \langle O_a(\vk_1) O_b(\vk_{23}) \rangle + 2\,\textrm{perm.} \right)+ B_{\epsilon,\unit}\, \diracpi(\vk_{123})\,, \nonumber
\ea
where the sum over contracted indices is implied.
The terms involving only $b_{O}$ coefficients give the deterministic contributions. The coefficients $P_{\epsilon,\unit}$ and $B_{\epsilon,\unit}$ give the constant stochastic contributions to the power spectrum and bispectrum, respectively, while the mixed $b_{O_a} P_{\epsilon,O_b}$ terms describe correlations between the deterministic galaxy density and the density-dependent stochastic variance. We apply the mean subtraction conditions $O(\vk=\bm{0})=0$, i.e. $\sum_{\vx}O(\vx)=0$ in \refeq{dd_general}.

\subsection{From the partition function to the renormalization group equations}\label{sec:RG}

To derive the RG equations we follow \cite{Rubira:2023vzw, rubira_renormalization_2024}, which employ the Wilson--Polchinski path-integral formulation \cite{Wilson:1971bg,POLCHINSKI1984269,Carroll:2013oxa}. An alternative derivation, based on the invariance of correlation functions of the galaxy density field on the cutoff choice, can be found in \cite{Bakx:2025cvu, bakx2026}.

We start by splitting the functional integral in \refeq{partition_function} into two parts, separated by cutoffs $\Lambda$ and $\Lambda'$,
\begin{equation}
    \delta_{\Lambda'}^{(1)}(\vk)=\dlinL(\vk)+\dshell(\vk),
\end{equation}
where $\dshell(\vk)$ is supported only in an infinitesimal shell $[\Lambda,\Lambda']$ of width $\lambda=\Lambda'-\Lambda$. After expanding $O[\dlinL(\vk)+\dshell(\vk)]$ in powers of $\dshell$ and integrating out the shell modes, we get a partition function with correction terms $\Zcorr{m}[J,\dlinL]$ (see \cite{rubira_renormalization_2024} for details):
\ba \label{eq:Zpluscorrections}
\Z[J] &= \int \Del\dlin_\L\, \P[\dlin_\L] \\
& \quad \times \exp\left(\sum_m \bigg\{\frac{1}{m!} \int_{\vx} \left[ \left( J(\vx) \right)^m \sum_O C_O^{\{m\}}({\L'})\, O[\dlin_\L](\vx)\right]  + \Zcorr{m}[J, \dlin_\L]\bigg\} \right) \,. \nonumber
\ea
At one-loop order (leading in $\lambda$), the $m=1$ correction is
\be\label{eq:zeta_correction}
\Zcorr{1}[J,\dlinL] = \int_{\vx} J(\vx) b_a(\Lambda')\, \mathcal{S}_{O_a}^2[\dlinL](\vx)\,, 
\ee
where the shell-operator contractions read 
\be \label{eq:SO2_term}
\mathcal{S}_{O_a}^2[\dlinL] = \lambda\, \frac{d\sigma^2_\Lambda}{d\Lambda}\bigg|_{\Lambda} s_{ab}^\mathrm{1L}\,O_b[\dlinL]+\mathcal{O}(\lambda^2)\,, 
\ee
with 
\be
\frac{d\sigma^2_\Lambda}{d\Lambda}\bigg|_{\Lambda} = \frac{\Lambda^2 \Plin}{2\pi^2}\,.
\ee
{One can extend the calculation to $m \geq 2$, leading to the renormalization of stochastic terms \cite{rubira_renormalization_2024}.}

The coefficients $s_{ab}^\mathrm{1L}$ are obtained by expanding the kernel $K_{O}(\vp,-\vp,\dots)$ in the hard-$p$ limit (i.e., $p$ much larger than the sum of all other arguments) and averaging over the direction of $\vp$ \cite{Bakx:2025cvu}:
\be \label{eq:biaskernelexpansionsinglehard}
  K_{O_a}^{(n+2)}({\bm k}_1,\dots,{\bm k}_n,{\bm p},-{\bm p})\bigg|^{p\to\infty}_{\text{av}_{\hat{\bf p}}} = \frac{2}{(n+2)(n+1)} s_{ab}^{\rm 1L}\, K_{O_b}^{(n)}({\bm k}_1,\dots,{\bm k}_n)\,,
\ee
where sums over repeated indices are implicit. 
We neglect higher-derivative corrections in \refeq{biaskernelexpansionsinglehard} since their effect on the running of leading bias operators is negligible on large scales \cite{bakx2026}.  For the running of dark-matter counterterms as a function of the cutoff, see \cite{Karandikar_2024,Peron2025}.

At one-loop level, the entries of $s_{ab}^{\rm 1L}$ vanish whenever ${\rm Ord}(O_a) \geq {\rm Ord}(O_b) + 2$. We truncate the bias basis at third-order [$n=3$ in \refeq{biaskernelexpansionsinglehard}], with the entries of the matrix $s_{ab}^{\rm 1L}$ collected in \reftab{sab1L}. We refer to this truncated matrix as $s_{ab}^{[n\leq3]}$ and comment more about the effects of truncating the RG at some perturbative order in \refsec{trunc_rg}.
The coefficients of $s_{ab}^{\rm 1L}$ are obtained through the rotation from the bias basis $\Pi$ from \cite{Bakx:2025cvu} to the Galilean basis. 
One of the reasons why we use the Galilean basis is the fact that entries of $s_{ab}^{\rm 1L}$ sourced by $\mathcal{G}_2$ and $\mathcal{G}_3$ are zero as a direct consequence of the non-renormalization theorem for Galilean operators \cite{assassi_renormalized_2014,Nicolis:2008in}, only contributing to higher-derivative terms. Further, $\Gamma_3$ is not sourced at third order.

\begin{table}[t]
\centering
\small
\renewcommand{\arraystretch}{1.25}

\begin{tabular}{lc|cc|cccc}
\toprule
$s_{ab}^{\mathrm{1L}}$
& $\delta$
& $\delta^2$
& $\mathcal{G}_2$
& $\delta^3$
& $\delta \mathcal{G}_2$
& $\mathcal{G}_3$
& $\Gamma_3$

\\
\midrule

$\delta$
& 0 & 0 & 0 & 0 & 0 & 0 & 0

\\
\midrule

\addlinespace[2pt]

$\delta^2$
& $\tfrac{68}{21}$
& $\tfrac{8126}{2205}$
& $\tfrac{254}{2205}$
& $\tfrac{286004}{169785}$
& $\tfrac{14458}{33957}$
& $\tfrac{1684}{101871}$
& $-\tfrac{661}{2695}$

\\

$\mathcal{G}_2$
& 0 & 0 & 0 & 0 & 0 & 0 & 0

\\
\midrule

\addlinespace[2pt]

$\delta^3$
& $3$
& $\tfrac{68}{7}$
& $0$
& $\tfrac{8126}{735}$
& $\tfrac{254}{735}$
& $0$
& $0$

\\

$\delta \mathcal{G}_2$
& $-\tfrac{4}{3}$
& $-\tfrac{376}{105}$
& $\tfrac{116}{105}$
& $-\tfrac{4628}{1715}$
& $\tfrac{18461}{5145}$
& $-\tfrac{3806}{15435}$
& $-\tfrac{125}{126}$

\\

$\mathcal{G}_3$
& 0 & 0 & 0 & 0 & 0 & 0 & 0

\\

$\Gamma_3$
& 0 & 0 & 0 & 0 & 0 & 0 & 0

\\

\bottomrule

\end{tabular}
\caption{Coefficients $s_{ab}^{\mathrm{1L}}$ in the Galilean basis. The index $a$ is the row and the index $b$ is the column.}
\label{tab:sab1L}
\end{table}

\begin{figure}
    \centering
    \includegraphics[width=0.5\linewidth]{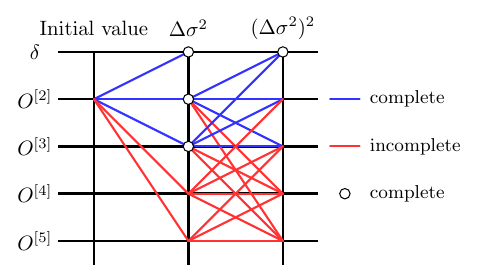}
    \caption{Diagram illustrating how the bias operators source each other at order $(\Ds)^2$. We assume initial values are zero for coefficients of order starting from 3. As for the linear bias, it only sources higher-derivative terms. An operator of order $n$ only sources other operators of order $n-2$ or higher at order $\Delta \sigma^2$. We truncate the bias basis at third order ($s_{ab}^{[n\leq3]}$), such that blue lines represent complete connections included in this work, while red, incomplete connections are not considered. The circles at the vertices show the final accuracy level for $O^{[\leq3]}$. }
    \label{fig:fullRG}
\end{figure}

We can relate the bias parameters at two cutoffs using \refeq{Zpluscorrections} to obtain the RG equations for $b_a(\Lambda)$ 
\be \label{eq:b_difftemplate}
b_a(\Lambda) = b_a(\Lambda') + \lambda \frac{d \sigma^2_\Lambda}{d \Lambda}\bigg|_{\Lambda}\, s_{ba}^\mathrm{1L}\, b_{b}(\Lambda')\,.
\ee
Taking the limit $\lambda\to 0$ yields the one-loop RG equation \cite{Rubira:2023vzw}
\be\label{eq:general_rg_equation}
\boxed{\frac{db_a}{d\Lambda}\bigg|_{\text{1L}} = -\frac{d\sigma^2_\Lambda}{d\Lambda}\, s_{ba}^\mathrm{1L}\, b_{b}(\Lambda).}
\ee
{Notice that the RGE of the bias \emph{coefficients} is described by minus the transpose of the matrix that describes the RGE of the \emph{operators}, corresponding to the differential (generator) version of the general bias transformation law (App. C of \cite{desjacques_large-scale_2018}).}
After changing variables to $\sigma^2$, the ODE admits an exponential solution, which can be expanded in powers of $\Delta\sigma^2=\sigma^2_*-\sigma^2$ 
\ba\label{eq:ODEsolution}
b_a\left(\sigma^2(\Lambda)\right) &= \left[e^{\bar{s}^{\rm 1L}\times\Ds}\right]_{ac}b_c^* \\
        &= b_a^* + \Ds\,\bar{s}_{ac}^{\rm 1L}\, b_c^* + \frac{1}{2}(\Ds)^2\,\bar{s}_{ab}^{\rm 1L}\bar{s}^{\rm 1L}_{bc}\, b_c^* + \frac{1}{6}(\Ds)^3\,\bar{s}^{\rm 1L}_{ab}\bar{s}_{bd}^{\rm 1L}\bar{s}^{\rm 1L}_{dc}\, b_c^* \vs
        & \hspace{9.5cm} + \mathcal{O}\!\left[(\Ds)^4\right]\,, \nonumber
\ea
where $\bar{s}^\mathrm{1L}_{ba} = s^\mathrm{1L}_{ab}$ denotes the transpose and starred quantities are evaluated at the initial scale $\Lambda_*$. Expanding the full RG solution makes clear the importance of treating the running of the bias parameters in full, since it resums the higher powers of $\sigma^2$ generated by higher-loop diagrams, in contrast to only retaining the leading $\sigma^2$ piece \cite{assassi_renormalized_2014}. Moreover, because the entries of $s_{ab}^{\rm 1L}$ vanish for ${\rm Ord}(O_a) > {\rm Ord}(O_b) + 2$, an operator of order $n$ contributes to one of order $n'<n$ only starting at order $(\Delta\sigma^2)^{\left\lceil\frac{n-n'}{2}\right\rceil}$. This pattern is illustrated in \reffig{fullRG}. 
Going beyond the current truncation plotted in \reffig{fullRG}, would require evaluating bias kernels up to sixth order.
The detailed truncation of the RG equations at a finite bias order is discussed in \refsec{trunc_rg}.

We now consider the two-loop contributions in the RG equations. Extending the Wilson--Polchinski derivation to two loops is non-trivial, as it requires path-integrating four momentum shells $\mathcal{S}_O^4$ over distinct support regions. The two-loop result was instead obtained in \cite{Bakx:2025cvu} using the independence of $n$-point functions of $\delta_g$ on the cutoff, which we reproduce here. The two-loop correction for $b_\delta$ is
\ba\label{eq:RG2Lb1}
 \frac{db_\delta}{d\Lambda}\bigg|_{\text{2L}} &= -30\, b_b\, \tilde d^{(5)}_b\, \frac{d\sigma^2_\Lambda}{d\Lambda}\int_0^\Lambda dq\, \frac{q^2\Plin(q)}{2\pi^2}\, g(q/\Lambda) \nonumber\\
 &=-30\, b_b\, \tilde d^{(5)}_b\, \frac{d\sigma^2_\Lambda}{d\Lambda}\int_0^\Lambda dq\, \frac{d\sigma^2(q)}{dq}\, g(q/\Lambda),
\ea
where $\tilde d^{(5)}_b$ sets the two-loop renormalization contribution of an operator $b_b$ into $\delta$ [see Eq. (B.1) in \cite{Bakx:2025cvu}] and
\begin{eqnarray}\label{eq:doublehardfuncg}
  g(r) = \frac{15r^8 - 40r^6 + 18r^4 - 40r^2 + 15}{240r^4}
  -\frac{(r^2-1)^4(r^2+1)}{64r^5}\ln\!\left(\frac{1+r}{1-r}\right)^2\,.
\end{eqnarray}
As discussed in \cite{Bakx:2025cvu}, the two-loop contribution is of the same order as the iterated one-loop (``one-loop squared'') contribution, i.e.\ an extra factor of $\sigma^2$ from the integral in \refeq{doublehardfuncg}. However, it is numerically suppressed for typical $\Lambda$CDM cosmologies; we find
\be\label{eq:doublehardfuncg_integral}
\int_0^\Lambda dq\, \frac{d\sigma^2(q)}{dq}\, g(q/\Lambda)\lesssim 0.12{\sigma^2(\Lambda)} \,,
\ee
such that we can neglect the contributions within the accuracy level required in this work.

{Similar expressions to \refeq{biaskernelexpansionsinglehard} in terms of hard limits involving multiple kernel products provide the renormalization for stochastic terms \cite{rubira_renormalization_2024,Bakx:2025cvu}. We leave the study of the running of stochastic contributions to future work.}

\subsection{Truncation of the RG}\label{sec:trunc_rg}

Comparing the RG equations for the bias parameters to lattice measurements involves two distinct truncations, one in the order of the bias expansion, and one in the order of perturbative gravity considered when evolving the initial conditions. We discuss each in turn.

First, the truncation of the bias order. The one-loop solution \refeq{ODEsolution} implies that an operator of order $n$ contributes to the running of an operator of order $n'<n$ starting at order $(\Delta\sigma^2)^{\left\lceil\frac{n-n'}{2}\right\rceil}$. Setting non-zero initial conditions only for the bias coefficients up to a fixed order is therefore not sufficient: higher-order operators are generated by lower-order ones at $\mathcal{O}(\Ds)$ (see \reffig{fullRG}) and subsequently back-react onto the lower-order parameters. A complete description of the RG would thus require computing $s^{\rm 1L}_{bc}$ for an infinite set of operators, including higher-loop contributions since the two-loop is only suppressed at $\mathcal{O}(\sigma^2(\Lambda)\Ds)$ in \refeq{RG2Lb1} and \refeq{doublehardfuncg_integral}.

This truncation of the RG, however, is a fundamental feature of perturbative theories, which are only meant to reproduce the underlying theory up to a given order in the expansion parameters, at the precision required by the problem. The RG therefore captures the correct running up to a fixed order in $\Ds$. At the time of writing, the matrix $s^{\rm 1L}_{bc}$ has been computed up to $n+2=5$ in \refeq{biaskernelexpansionsinglehard} \cite{Bakx:2025cvu}, i.e.\ ${\rm Ord}(O_b) \leq 5$ and ${\rm Ord}(O_c) \leq 3$.  Setting the initial conditions of bias coefficients of order $\geq 3$ to zero, we truncate the RG matrix $s^{\rm 1L}_{bc}$ at ${\rm Ord}(c) \leq 3$ and ${\rm Ord}(b) \leq 3$,  yielding (see \reffig{fullRG}):
\begin{itemize}
    \item a prediction for $b_\delta$ at $\mathcal{O}((\Ds)^2)$;
    \item a prediction for $b_{O^{[2]}}$ at $\mathcal{O}(\Ds)$;
\end{itemize}

We now discuss the truncation in terms of the gravity order in Eulerian space. The structure of the $n$-shell-operator contractions \refeq{SO2_term} is set by limits of the kernels $K^{(n)}_O$ in \refeq{biaskernelexpansionsinglehard}, which encode the forward model relating initial conditions to the final positions. In practice, any such forward model is itself truncated at some order in the computation of the matter density field.
Considering the construction of the kernels $K^{(n)}_O$ for the bias operators $\delta^2$, we have
\ba\label{eq:bias_rg_relation_matter_kernel}
K_{\delta^2}^{(3)}({\bm k},{\bm p},-{\bm p}) &\propto 2F_2(\vk,\vp)\,, \nonumber\\
K_{\delta^2}^{(4)}({\bm k}_1,{\bm k}_2,{\bm p},-{\bm p}) &\propto \left[6F_3(\vp,\vk_1,\vk_2)+4F_2(\vp,\vk_1)F_2(-\vp,\vk_2)\right]\,, \\
K_{\delta^2}^{(5)}({\bm k}_1,{\bm k}_2,{\bm k}_3,{\bm p},-{\bm p}) &\propto \left[12 F_2(\vp,\vk_1)F_3(-\vp,\vk_2,\vk_3)+8 F_4(\vp,\vk_1,\vk_2,\vk_3)\right]\,,\nonumber 
\ea 
with $F_n$ the standard perturbation theory (SPT) kernels \cite{bernardeau_large-scale_2002} and
where the part of the kernels sourcing to higher-derivative terms, such as $k^2 \dlinL(\vk)$ are ignored above. The first line generates the contribution of $\delta^2$ to $\delta$, the second sources $b_{O^{[2]}}$, and the third sources $b_{O^{[3]}}$. Truncating gravity at $F_{2}$ thus captures the running of $\delta$ from $\delta^2$ correctly, while giving only partial contributions to second-order operators. Obtaining the full running of $b_{O^{[3]}}$ induced by $b_{\delta^2}$ requires expanding the second-order bias kernel up to fourth order in perturbation theory, including the full $F_{4}$ SPT kernel. The same argument can be generalized to the sourcing from higher-order operators.

\begin{figure}
    \centering
    \includegraphics[width=1.0\linewidth]{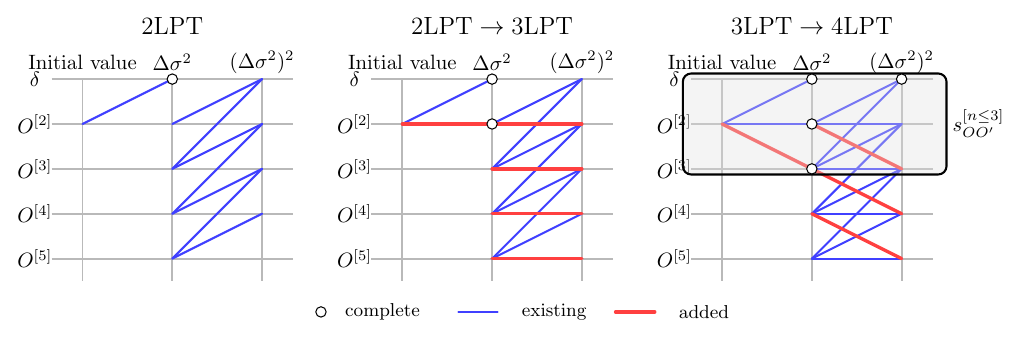}
    \caption{Sourcing diagrams between different orders of bias operators for 2LPT, 3LPT and 4LPT matter forward models. The sourcing at a specific $(\Ds)^n$ is complete only if all the connection lines linking back to the starting point at initial conditions are complete. The shaded area in the 4LPT plot represents the matching between theoretical prediction with the sourcing matrix $s_{ab}^{[n\leq3]}$ truncated at third order. The lines `existing' and `added' indicate, respectively, the contributions present in $n$LPT and those introduced when increasing the forward-model order to $(n+1)$LPT. }
    \label{fig:sourcing_diagram_1L}
\end{figure}

Since \texttt{LEFTfield} forward-models the initial conditions using Lagrangian displacements (see \refsec{forward_model}), it is also important to consider how the truncation of LPT displacements affects the RG equations above, derived in Eulerian space. The SPT kernels can be decomposed in terms of LPT kernels as \cite{Rampf_2012b}
\be
F_n \supset \prod_{n_i;n_1 + \dots + n_j = n} \vk \cdot\bm{S}^{(n_i)}_{L\oplus T} \,,
\ee
where $\bm{S}^{(n)}_{L\oplus T}$ is the shorthand notation for $n^{\rm th}$-order LPT symmetrized longitudinal and transverse kernels, see the explicit definition in \refeq{LPT_kernel_def_L} and \refeq{LPT_kernel_def_T}. The complete relation up to fourth-order SPT kernels is given in \refapp{LPT_order} and we discuss LPT further in \refsec{forward_model}. Reproducing the correct SPT kernel $F_{n}$, and hence the RG running encoded in $s^{\rm 1L}_{bc}$, therefore requires going to at least the same order in LPT, while higher-order SPT kernels are partially generated. The diagrams in \reffig{sourcing_diagram_1L} illustrate the completeness of the SPT RG structure at a truncated LPT order: red lines indicate incomplete vertex structures, blue lines complete ones. At 2LPT and 3LPT the running of the linear bias is correct only at $\mathcal{O}(\Ds)$, while at 4LPT it is complete up to $\mathcal{O}((\Ds)^2)$. The running of quadratic operators is complete at $\mathcal{O}(\Ds)$ only from 3LPT onward. We discuss the convergence under the LPT perturbative order in \refapp{nLPT} .

%%%%%%%%%%%%%%%%%%%%%%%%%%%%%%%%%%%%%%%%%%%%%%%%%%%%%%%%%%%%%%%%%%%%%%%%%%%
%%%%%%%%%%%%%%%%%%%%%%%%%%%%%%%%%%%%%%%%%%%%%%%%%%%%%%%%%%%%%%%%%%%%%%%%%%%

\section{EFT forward model, simulations, and bias measurement methodology}\label{sec:simulation_measurement}

In this section, we describe our forward model for the galaxy density field $\dg$ and its dependence on the filtered initial conditions $\dlinL$. We then introduce the mock and halo data and the field-level likelihood used to infer the bias coefficients. Finally, we describe how we measure these coefficients using traditional $n$-point correlation functions.

\subsection{Forward-modeling biased tracers}\label{sec:forward_model}

The reference galaxy density field $\hdg$ and the sampled galaxy density field $\dg$ used for the running measurement are both generated with \texttt{LEFTfield} \cite{schmidt_rigorous_2018}, a Lagrangian perturbation theory forward model for matter evolution, with biased tracers subsequently constructed from operators in the Eulerian frame.
We discuss below finite lattice-resolution artifacts and how the perturbative order, such as the LPT and bias-expansion orders, are chosen to achieve the required accuracy up to a fixed scale, to be later compared with the RG running.

The components of the forward model are (see \cite{schmidt_rigorous_2018,stadler_fast_2025} for more details):
\begin{enumerate}
\item {\bf Initial conditions:} The linear density field used as initial conditions $\dlinL$ is filtered at a cutoff scale $\Lambda$ with a top-hat filter $W_\Lambda(k)$ in Fourier space, to keep modes perturbatively controlled \cite{schmidt_unbiased_2020}. The initial field is created by a 3D Gaussian random realization $\hat{s}(\vk)$ drawn from $\mathcal{N}(0,1)$ at a given box size $L$. The grid size $N_{\rm G,Ny}$ required to represent a given mode $k_\star$, such that $k_{\rm Ny}(N_{\rm G,Ny}) > k_\star$, is given by
    \be \label{eq:NNy}
    N_\mathrm{G,Ny}(k_\star)=\Bigl\lceil \frac{k_\star L}{\pi}\Bigr\rceil\,.
    \ee
    We then define the grid size $N_\mathrm{G,in} \equiv N_\mathrm{G,Ny}(\Lambda)$ as the smallest integer such that its Nyquist frequency exceeds $\Lambda$, the initial condition filtering.
    Therefore, for a given $\Lambda$ and linear power spectrum $\Plin(k,\tau_0)$ at the present time $\tau_0$
    \begin{equation}
        \delta_\Lambda^{(1)}(\vk,\tau_0)=W_\Lambda(k)\sqrt{L^3 \Plin(k,\tau_0)}\hat{s}(\vk)\,.
    \end{equation}
    
    \item {\bf Evolution:} The gravitational evolution described by LPT relates the particle's initial position $\vq$ to its final position $\vx$ as follows
    \begin{equation}
        \vx(\vq,\tau)=\vq+\sum_{n=1}^{n_\mathrm{LPT}}\bm{s}^{(n)}(\vq,\tau)\,,
    \end{equation}
    where the perturbative displacement $\bm{s}^{(n)}$ is determined by the Poisson equation and the displacement field equation of motion \cite{schmidt_n-th_2021}. Here $n_\mathrm{LPT}$ is the LPT order considered.
    
    At each order, $\bm{s}^{(n)}$ is decomposed into a longitudinal component $S^{(n)}_L$ and a transverse component $\bm{S}^{(n)}_T$, see \refapp{LPT_order}, which are obtained iteratively from recursion relations involving the Lagrangian distortion tensor $H_{ij}=\frac{\partial}{\partial q_i} s_j(\vq,\tau)$ \cite{schmidt_n-th_2021}.
    The time evolution is factorized out following the discussion in \refsec{introduction}.
    The grid size for the forward evolution is determined by $N_\mathrm{G,fwd} = N_\mathrm{G,Ny}(n_\mathrm{LPT} \Lambda)$ using \refeq{NNy}, such that all modes generated by the perturbative kernels are resolved.
    
    \item {\bf The final density field:} Particles are then assigned to an Eulerian grid of size $N_{\rm G,Eul}$ to obtain the evolved matter density contrast from the Lagrangian coordinates $\vq$. From mass conservation
    \begin{equation}
        1+\delta_\mathrm{fwd}(\vx(\vq),\tau_0)=\left| \frac{\partial \vx}{\partial \vq}\right|^{-1}_{\vx=\vx(\vq, \tau_0)}\,,
    \end{equation}
    or, in Fourier space,
    \be\label{eq:density_assignment_fourier}
    \delta_\mathrm{fwd}(\vk,\tau_0)=\int d^3q e^{-i\vk\cdot\vq}[e^{-i\vk\cdot\bm{s}(\vq,\tau_0)}-1]\,.
    \ee
    We use the NUFFT scheme \cite{barnett2019parallelnonuniformfastfourier} for the density assignment.
    The operation in \refeq{density_assignment_fourier} is fully nonlinear. Mass is conserved fully nonlinearly, but the finite grid introduces truncation errors \cite{stadler_fast_2025}.
    We take $N_\mathrm{G,Eul}=N_\mathrm{G,Ny}(4 \Lambda)$, so that the $F_4$ kernel in \refeq{bias_rg_relation_matter_kernel} is not truncated. This allows
    for a consistent sourcing of $O^{[2]}$ to $O^{[3]}$, as required by \refeq{bias_rg_relation_matter_kernel}. We discuss the effect of changing $N_\mathrm{G,Eul}$ in \refapp{resolution_effect}.
    
    \item {\bf Operators:} The bias operators $O$ and the deterministic large-scale galaxy density are then constructed from the evolved matter density contrast $\delta_\mathrm{fwd}$ as
    \begin{equation}
        \delta_{g,\mathrm{det}}(\vk)=\sum_{O}b_O O[\delta_\mathrm{fwd}](\vk)\,.
    \end{equation}
    The prediction for modes $\vk$ in $\dg$ is accurate only for $|\vk|<\Lambda$. 
    We focus on the running of leading-in-gradient operators, so the running of higher-derivative corrections is neglected, but we include $\nabla^2 \delta$ when sampling the bias parameters.    
    Therefore, our list of operators up to ${\rm Ord}(O)=3$ is
    \be
    O\in \{ \delta,\delta^2,\mathcal{G}_2,\delta^3,\delta \mathcal{G}_2, \mathcal{G}_3, \Gamma_3,\lapl\delta\}.
    \ee
    The operators $O$ are built on a grid size $N_\mathrm{G,final}=n_\mathrm{bias}n_\mathrm{LPT}\Lambda$, where $n_\mathrm{bias}$ is the order of the bias expansion considered. Lastly, the comparison with the data $\delta_g$ is performed for modes with $|\vk| < k_\mathrm{max}$ using the likelihood discussed in \refsec{field_measure}, such that the grid size used in the likelihood is $N_\mathrm{G,LH}=N_\mathrm{G,Ny}(k_\mathrm{max})$.
\end{enumerate}

\begin{figure}
    \centering
    \includegraphics[width=1.\linewidth]{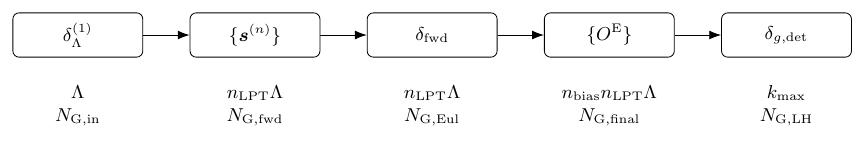}
    \caption{A summary of the choices of grid sizes and their $k$-mode information at each step of the \texttt{LEFTfield} pipeline employed in this work. The corresponding grid size $N_\mathrm{G,i}$ is determined by plugging the filter value $\Lambda_i$ into \refeq{NNy}.}
    \label{fig:LEFTfield}
\end{figure}

Stochasticity is also required to account for uncertainties from small-scale physics when constructing the galaxy density field  \cite{rubira_renormalization_2024,rubira2025nongaussiangalaxystochasticitynoisefield}. We consider the simplest case, with a single Gaussian noise field $\epsilon$ added to the deterministic galaxy density field $\dgdet$, which is marginalized within the likelihood we describe in \refsec{field_measure}.
We fix all cosmological parameters to $\{ \Omega_m=0.3, \Omega_\Lambda=0.7,h=0.7, n_s=0.967,\sigma_8=0.85\}$ for the mocks and samples in this work.
Throughout, the forward-modeled volume is a cubic box with sidelength $L=2000\ \mathrm{Mpc}/h$ and periodic boundary conditions. The grid size $N_\mathrm{G}$ is chosen for each value of $\Lambda$, following the pipeline described above.
\reffig{LEFTfield} summarizes the grid size choices in each step of the mock generation.

\subsection{Data} \label{sec:data}

We consider the running of bias parameters for two different kinds of data:
\begin{enumerate}
    \item Mocks generated by \texttt{LEFTfield} (results in \refsec{res_mocks}). \\
    A suite of 100 mocks is generated with $(b^{\Lambda_*}_\delta, b^{\Lambda_*}_{\delta^2}, b^{\Lambda_*}_{\mathcal{G}_2},\sigma_\epsilon^{\Lambda_*},\sigma_{\epsilon,2}^{\Lambda_*})=(2,1,0, 0.2, 0)$, and all higher-order bias parameters set to zero, at the cutoff ${\Lambda_*}=0.1\ h/\mathrm{Mpc}$ following the forward model introduced in \refsec{forward_model}. For reference, this corresponds to ${\sigma^2_*} \approx 0.16$ for the $\Lambda$CDM model considered. The noise parameters $\sigma_\epsilon^{\Lambda_*}$ and $\sigma_{\epsilon,2}^{\Lambda_*}$ enter the likelihood and are defined below, in \refsec{field_measure}. Since the input bias parameters at $\Lambda_*$ are known, these mocks provide a controlled setting for studying the running of bias parameters.
    
    \item Dark matter halo catalog (results in \refsec{res_halos}). \\
    This single catalog is based on the N-body, gravity-only simulations presented in \cite{Biagetti_2017}, which have also been used in \cite{Schmidt_2021}. It was generated using GADGET-2 \cite{springel_2001,Springel_2005} for a flat $\Lambda$CDM cosmology with parameters and a box size same as the forward model in \refsec{forward_model}, and $1536^3$ dark matter particles of mass $M_\mathrm{part}=1.8\times10^{11}\ h^{-1}M_\odot$. Initial conditions are generated at redshift $z_\mathrm{ini}=99$ using 2LPT with the 2LPTic algorithm \cite{Crocce_2006,Scoccimarro_2012}. Then dark matter halos are identified at redshifts $z=0,0.5,$ and 1 with the \texttt{ROCKSTAR} halo finder \cite{Behroozi_2012} in a set of mass ranges $\log_{10} M_{200m} = 12.5\text{-}13.0, 13.0\text{-}13.5, 13.5\text{-}14.0, 14.0\text{-}14.5\ h^{-1}M_\odot$ mass ranges, with an overdensity threshold of 200 times the background matter density. 
\end{enumerate}

\subsection{Field-level inference of bias parameters}\label{sec:field_measure}

Inferring bias parameters [or, more generally, $C^{\{m\}}_O$ from \refeq{effective_action}] from field-level data requires a likelihood $\mathcal{L}(\dg |{C^{\{m\}}_O}, \dlin)$.
From this likelihood, one can infer the posterior $\mathcal{P}({C^{\{m\}}_O}, \dlin|\dg)$ in a full Bayesian framework, adopting priors for $\dlinL$ and ${C^{\{m\},\Lambda}_O}$ and sampling it using galaxy-density realizations $\dg=\sum_O b_O^{\Lambda}O[\delta^{(1)}_{\Lambda}]$ evaluated at a lower cutoff $\Lambda<\Lambda_*$ \cite{schmidt_rigorous_2018}.
As shown in \cite{rubira2025nongaussiangalaxystochasticitynoisefield}, considering the full set of coefficients $C^{\{m\}}_O$ from \refeq{effective_action} is equivalent to sampling an extra field $\epsilon$, which makes the likelihood sampling very expensive. In this paper, we consider instead the \textit{Gaussian} EFT likelihood $\mathcal{L}$ (up to a constant factor) 
\ba\label{eq:likelihood}
\ln\mathcal{L}[\hdgkmax|\dlinL,\{b_O\},\sigma_\epsilon,\sigma_{\epsilon,2}]&=\vs\
&\hspace{-2cm}-\frac{1}{2}\sum_{\vk\neq0}^{\kmax}\left[\ln 2\pi P_\epsilon(k)+\frac{1}{P_\epsilon(k)}\big| \hdg(\vk)-\dgdet[\{b_O\},\dlinL](\vk) \big|^2\right] \,,
\ea
where $\kmax$ determines the $\vk$ modes used in the inference.  We fix $\kmax$ for the inference at different $\Lambda$. Also $P_{\epsilon}(k)$ is the power spectrum of the Gaussian noise field $\epsilon$, which as mentioned in \refsec{RG} also runs with the cutoff $\Lambda$ \cite{rubira_renormalization_2024}, and is sampled together with the subleading scale-dependent $k^2$ correction,
\be\label{eq:noise_amplitude}
P_\epsilon(k) =\frac{L^3}{N_\mathrm{G}^3} \sigma^2_\epsilon (1+\sigma_{\epsilon,2}k^2)^2 \,.
\ee
One might suspect that the incomplete noise modeling in the likelihood may bias the inference, since the non-Gaussian or density-dependent terms such as $P_{\epsilon,\delta}$ are generated by the running. However, we show in \refapp{FLI_bias_estimation} that \refeq{likelihood} still provides an unbiased estimation under the ensemble of realizations, and merely the uncertainty estimate is affected by the Gaussian likelihood assumption at leading order.

In this work, we focus on the running of the first- and second-order bias parameters ${b_\delta,b_{\delta^2},b_{\mathcal{G}2}}$. We therefore explicitly sample the operators $O_{\mathrm{sampled}} = {O_\delta,O_{\delta^2},O_{\mathcal{G}2}}$. Instead of sampling the higher-order parameters directly, which increases both the dimensionality of the parameter space and the length of the sampling chains, we analytically marginalize over $b_{O^{[3]}}$. 
From the EFT likelihood in \refeq{likelihood}, we marginalize over $O_\mathrm{marg} = {O^{[3]}}$ assuming a Gaussian prior $\mathcal{N}(\bar{\mathbf{b}},\mathbf{\Sigma}_b)$, with mean $\bar{\mathbf{b}}$ and covariance $\mathbf{\Sigma}_b$, obtaining \cite{elsner_cosmology_2020,Kosti_2023}
\ba\label{eq:marginalized_likelihood}
\ln\mathcal{L}[\hdgkmax|\dinL,\sigma_\epsilon,\sigma_{\epsilon,2}]=&-\frac{1}{2}\sum_{\vk\neq0}^{\kmax}\left[\ln 2\pi P_\epsilon(k)+\frac{1}{P_\epsilon(k)}\big| \hdg(\vk)-\sum_{O \subset O_\mathrm{sampled}}b_OO(\vk)] \big|^2\right] \nonumber\\  
&\hspace{-2cm}-\frac{1}{2}\left(\tr\ln|\mathbf{\Sigma}_b| +\tr\ln |\mathbf{A}'|-\mathbf{B}'^\dagger\mathbf{A}'^{-1}\mathbf{B}'+\bar{\mathbf{b}}^T\mathbf{\Sigma}_b^{-1}\bar{\mathbf{b}}\right)\,,
\ea
where 
\ba 
\mathbf{A'}&=\mathbf{A}+\mathbf{\Sigma}_b^{-1}, \quad \mathrm{with}\ A_{O_\mathrm{marg}O_\mathrm{marg}'}=\sum_{\vk\neq0}^{\kmax} \frac{O_\mathrm{marg}(\vk)O_\mathrm{marg}'^*(\vk)}{P_\epsilon(\vk)}\,,\label{eq:marginalized_A} \\
\mathbf{B}'&=\mathbf{B}+\mathbf{\Sigma}_b^{-1}\bar{\mathbf{b}}, \quad\mathrm{with}\ B_{O_\mathrm{marg}}=\sum_{\vk\neq0}^{\kmax}\frac{[\hdg(k)-\sum_{O\subset O_\mathrm{sampled}}b_OO(\vk)]O_\mathrm{marg}^*(\vk)}{P_\epsilon(\vk)}\,.\label{eq:marginalized_B}
\ea
This likelihood marginalizes over the effect of the unsampled leading higher-order bias parameters. 
Note that a likelihood in which bias parameters of order higher than $n$ are marginalized is not equivalent to a likelihood in which the bias expansion is simply truncated at $n$th order (see \refapp{marginalization}).

For mocks and halo samples described in \refsec{data}, we fix $k_\mathrm{max}$ in \refeq{marginalized_likelihood} and ensure $\kmax\ll\Lambda$ for each sampled mock generated at $\Lambda<\Lambda_*$. Our target posterior is
\begin{equation}\label{eq}
\mathcal{P}({b_\delta^\Lambda, b_{\delta^2}^\Lambda, b_{\mathcal{G}2}^\Lambda, b_{\lapl}^\Lambda}, {\sigma_\epsilon^\Lambda, \sigma^\Lambda_{\epsilon,2}}|{\hdgkmax}).
\end{equation}
For mocks generated by a 4LPT matter forward model, we expect the theoretical prediction based on the mixing matrix $s_{ab}^{[n\leq3]}$ to describe the running of the third-order bias parameters accurately to leading order in $\Ds$ (see \reffig{sourcing_diagram_1L}). We therefore take the prediction from \refeq{ODEsolution} as the mean $\bar{\mathbf b}$ of the Gaussian prior.
We take $\mathbf{\Sigma}_b$ to be diagonal with a conservative unit variance for each marginalized parameter, corresponding to the expectation $b_{O^{[3]}}^\Lambda\sim\mathcal{O}(1)$ (see discussion in \refapp{marginalization}, which shows that this prior $\mathbf{\Sigma}_b$ does not significantly affect the inferred bias parameters). As for halo samples, since there is no prior knowledge on the values of higher-order bias parameters, we take $\mathbf{\Sigma}_b$ with the values of diagonal elements to be 100.

\begin{figure}
    \centering
    \includegraphics[width=0.8\linewidth]{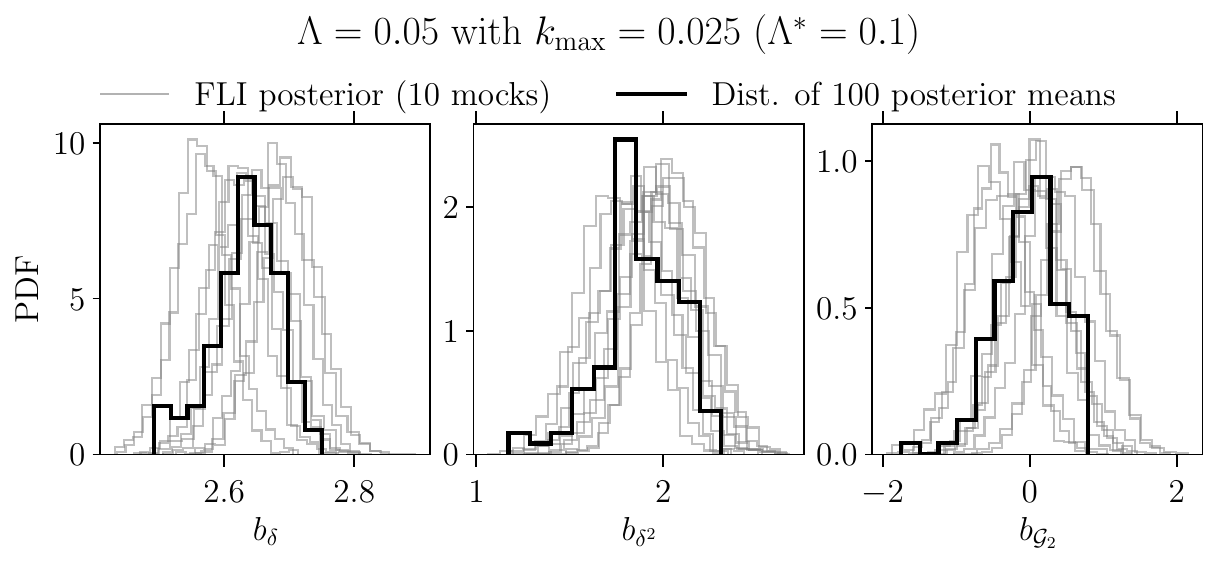}
    \caption{The histogram of 10 posteriors of the fixed-phase field-level inference sampling and the histogram of the posterior mean of 100 mocks. The values of $\kmax, \L$ are indicated in the figure. The mocks are generated at $\Lambda_*=0.1\ h/\mathrm{Mpc}$. }
    \label{fig:FLI_errorbar}
\end{figure}

The field-level inference considered in the work is conditioned on the known initial conditions field $\dlin(\vk)$. Matching the shell integration description in the partition function \refeq{Zpluscorrections}, it means that $\dlinL(\vk)$ is the same between the mocks and the samples, and the missing $\dshell$ information becomes the source of stochasticity and leads to realization-dependent $b_O(\Lambda)$. 
Due to the Gaussian likelihood approximation, the sample variance can in general be larger than that inferred from the posterior. \reffig{FLI_errorbar} shows the posterior of the means across different realizations compared to individual posteriors, where the former is around $10\%$ wider than the latter. The $b_{\delta^2}$ and $b_{\mathcal{G}_2}$ posteriors display mild non-Gaussianity across the 100 mocks considered.
For the \texttt{LEFTfield} mocks, which are easily created, we consider the ensemble estimate of the bias parameters from the $N_{\rm mock}=100$ independent realizations. We use the mean of the per-realization posterior means $\bar{b_{O}}_i$ as the inference result, and quote its uncertainty as
$\sigma_{\mathrm{FLI},\bar{b_O}} = \sigma_\mathrm{scatter}/\sqrt{N_{\rm mock}}$, where $\sigma_\mathrm{scatter}^2 = (N_{\rm mock}-1)^{-1}
\sum_i (\bar{b_{O}}_i-\bar{b_{O}})^2$ is the sample variance of the $\bar{b_{O}}_i$ across realizations. Notice that the sample variance captures all sources of variance induced by the fluctuating initial conditions. 
For the halo samples, we quote the mean of the posterior and the posterior variance as the uncertainty since there is only one realization. Given the above discussion, the bias uncertainties are likely to be underestimated.

\subsection{$n$-point function inference of bias parameters}\label{sec:npt}
We can write the deterministic part of the bias expansion either in terms of finite cutoff or in terms of renormalized bias operators $[O]$ and bias coefficients $b_{O}^{\npt}$
\begin{equation}\label{eq:bias_expansion_cutoff_dependence_npt}
    \delta_g^{\rm det}(\vx,\tau)=\sum_O b_O^\Lambda(\tau)O(\vx,\tau)=\sum_O b_{O}^{\npt}(\tau)[O](\vx,\tau),
\end{equation}
in which the renormalized operators are defined through the \textit{renormalization conditions}, \refeq{renormalization_conditions}.
The coefficients $b_{O}^{\npt}$ used in $n$-pt function measurements should be equivalent to the field-level bias parameters in the limit $\Lambda\to 0$ \cite{rubira_renormalization_2024}. 

Our aim here is to measure the linear and second-order $b_O^{\npt}$ using the tree-level power spectrum and bispectrum. Motivated by \refeq{renormalization_conditions} (with $n=1,2$), and since we have the initial conditions available for both mocks and halos, we consider the cross-correlation of the tracer with \emph{linear} density fields:
\ba\label{eq:cross_spectrum}
{P}_{gl}(k) &\equiv \langle\delta_g(\vk) \dlin(-\vk)\rangle' = \left[b_{\delta}^{\npt} - b_{\lapl\delta}^{\npt} k^2 \right]\Plin(k) \,, \\
{B}_{gll}(k_1,k_2,k_3) &\equiv \langle\delta_g(\vk_1) \dlin(\vk_2)\dlin(\vk_3)\rangle' \nonumber\\
&=\left[2 b_{\delta}^{\npt} F_2(\bm{k}_2,\bm{k}_3) + 2b_{\delta^2}^{\npt} + 2b_{\mathcal{G}_2}^{\npt} \left( [\hat{\bm{k}}_2\cdot\hat{\bm{k}}_3]^2-1\right)- 2b_{\lapl\delta}^{\npt} k_1^2 F_2(\bm{k}_2,\bm{k}_3)\right] \nonumber\\
&\qquad \times\Plin(k_2)\Plin(k_3) \,.\label{eq:cross_bispectrum}
\ea 
We further include the leading-order higher-derivative terms $b_{\lapl\delta}$ to partially absorb higher-order corrections.
We estimate the bias parameters $\{b_\delta, b_{\delta^2}, b_{\mathcal{G}_2}, b_{\lapl\delta} \}$ given the measured cross-power spectrum $\hat{P}_{gl}(k)$, bispectrum $\hat{B}_{gll}(k_1,k_2,k_3)$ and the covariance matrix $\bm{C}$ estimated from mocks.
Specifically, we use binned estimators for the cross-power spectrum $\hat{P}_{gl}(k)$ and bispectrum $\hat{B}_{gll}(k_1,k_2,k_3)$: \cite{Oddo_2020,Oddo_2021}
\ba 
\hat{P}_{gl}(k)&=\frac{1}{N_P(k)}\sum_{|\bm{n}|\in k} P_{gl}(\bm{n}), \label{eq:P_gi_estimator} \\
\hat{B}_{gll}(k_1,k_2,k_3)&=\frac{1}{N_B(k_1,k_2,k_3)}\sum_{|\bm{n}_1|\in k_1}\sum_{|\bm{n}_2|\in k_2}\sum_{|\bm{n}_3|\in k_3}\diracpi(\bm{n}_{123})B_{gll}(\bm{n}_1,\bm{n}_2,\bm{n}_3), \label{eq:B_gii_estimator}
\ea 
where $N_P(k)$ and $N_B(k_1,k_2,k_3)$ are the numbers of the fundamental configurations within the bin range, and $|\bm{n}|\in k$ represents the grid whose magnitude $|\bm{n}|k_\mathrm{F}$ is located in the bin range $[k-\Delta k/2, k+\Delta k/2]$. We use linear bins with bin size equal to the fundamental mode of the simulation box, i.e. $\Delta k=k_\mathrm{F}$, which corresponds to 32 bins spaced linearly in $k$ with $k_\mathrm{max}=0.1\ h/\mathrm{Mpc}$. The bispectrum in the triangle bins with closed triangles is measured with the same linearly spaced $k$ sides.
In total, the data vector size is  $N_{P+B}\sim3800$.

For the mocks, we use the general Gaussian likelihood function $\mathcal{L}_{n\text{-pt}}$, 
\be\label{eq:npt_likelihood}
\ln\mathcal{L}_{n\text{-pt}}=-\frac{1}{2}(\hat{\bm{X}}-\bm{X}_\mathrm{model})\bm{C}^{-1}(\hat{\bm{X}}-\bm{X}_\mathrm{model})^T,
\ee 
where $\bm{X}=(P_{gl},B_{gll})$ is the vector spanned by all bins of $P_{gl}$ and $B_{gll}$. In practice, the mean of $P_{gl}$ and $B_{gll}$ over 100 mocks are taken as the data vector $\hat{\bm{X}}$. 
Data vectors from 100,000 mock realizations are used to estimate the precision matrix $\bm{C}^{-1}$, or precision matrix. This number ensures that the correction from the Hartlap factor \cite{Hartlap_2006} rescaling the estimated precision matrix is at most $4\%$ when using all $k$- and triangle-bins from $P_{gl}(k)$ and $B_{gll}(k_1,k_2,k_3)$.
The marginalized uncertainties of the parameters are estimated by the curvature around the maximum-likelihood point,
\be\label{eq:ml_uncertainty}
\sigma(\theta_i)=\sqrt{\left(\frac{\partial^2\mathcal{L}_{\npt}}{\partial\bm{\theta}\partial\bm{\theta}}\right)^{-1}_{ii}},
\ee 
where $\theta_i$ are the bias parameters to be fit in the model.
The error bars are reduced by a factor $1/\sqrt{N_\mathrm{mock}}$, since they quantify the uncertainty in the mean recovered value across the mock ensemble.

In the case of the dark matter halo catalog, for which the covariance is not available given that there is only one catalog, we compute the ratio of the cross power spectrum and bispectrum over the auto-spectrum $P_{ll}$ of the linear density field to suppress the cosmic variance \citep{Lazeyras:2015lgp},
\ba
r_{P_{gl}}(k)&=\frac{P_{gl}(k)}{P_{ll}(k)}, \\
r_{B_{gll}}(k_1,k_2,k_3)&=\frac{B_{gll}(k_1,k_2,k_3)}{(P_{ll}(k_2)P_{ll}(k_3))},
\ea
and fit the parameters with least-squares fitting. Since we do not have a covariance for $r_{P_{gl}}, r_{B_{gll}}$ available, we refrain from quoting error bars on the resulting bias coefficients. 
However, we have verified that the $n$-point function results are consistent with the bias relations calibrated from previous measurements using the separate universe approach and theoretical predictions \cite{Lazeyras:2015lgp,lazeyras_assembly_2021}, as described in \refapp{halo} (\reffig{halo_bias_relation}). This comparison indicates that the precision is at the level of a few percent or better.

For the $n$-point function measurement in the mocks described in \refsec{data}, we fit the measured cross-spectrum and bispectrum with $k_{P,\mathrm{max}}=0.1\ h/\mathrm{Mpc}$ and $k_{B,\mathrm{max}}=0.1\ h/\mathrm{Mpc}$, where no significant drift in the best-fit parameters due to the model incompleteness in \refeq{cross_spectrum} and \refeq{cross_bispectrum} up to the available $k$ modes is found. 
For the $n$-point function measurements in the halo catalog described in \refsec{data}, we fit with $k_{P,\mathrm{max}}=0.2\ h/\mathrm{Mpc}$ and $k_{B,\mathrm{max}}=0.2\ h/\mathrm{Mpc}$ in the least-squares fitting between $\hat{r}_{P_{gl}}$, $\hat{r}_{B_{gll}}$ and their bias model.

%%%%%%%%%%%%%%%%%%%%%%%%%%%%%%%%%%%%%%%%%%%%%%%%%%%%%%%%%%%%%%%%%%%%%%%%%%%
%%%%%%%%%%%%%%%%%%%%%%%%%%%%%%%%%%%%%%%%%%%%%%%%%%%%%%%%%%%%%%%%%%%%%%%%%%%
\section{Results}\label{sec:results}
In this section, we present results of the bias parameters measured at different cutoff scales $\Lambda$ using the field-level inference described in \refsec{field_measure} and the $n$-point function measurements of \refsec{npt}, applied to the data described in \refsec{data}.
In \refsec{res_mocks}, we describe the running of the bias parameters using mocks generated by the perturbative forward model, for which the bias parameters are known at the input cutoff $\Lambda_*$.
We then consider the bias running for dark matter halo catalogs in \refsec{res_halos}, which represent a more realistic scenario.
We sample $\{b_\delta^{\Lambda}, b_{\delta^2}^{\Lambda}, b_{\mathcal{G}_2}^{\Lambda}, b_{\lapl}^{\Lambda}, \sigma_\epsilon^\Lambda, \sigma^\Lambda_{\epsilon, k^2}\}$ using the likelihood described in \refsec{field_measure}, in which the third-order bias parameters are marginalized over.
In each case, we compare the results with the theoretical predictions obtained from the RG solutions of \refsec{theory}. 

\subsection{The running of bias parameters on mocks} \label{sec:res_mocks}

\begin{figure}
    \centering
    \includegraphics[width=0.8
    \linewidth]{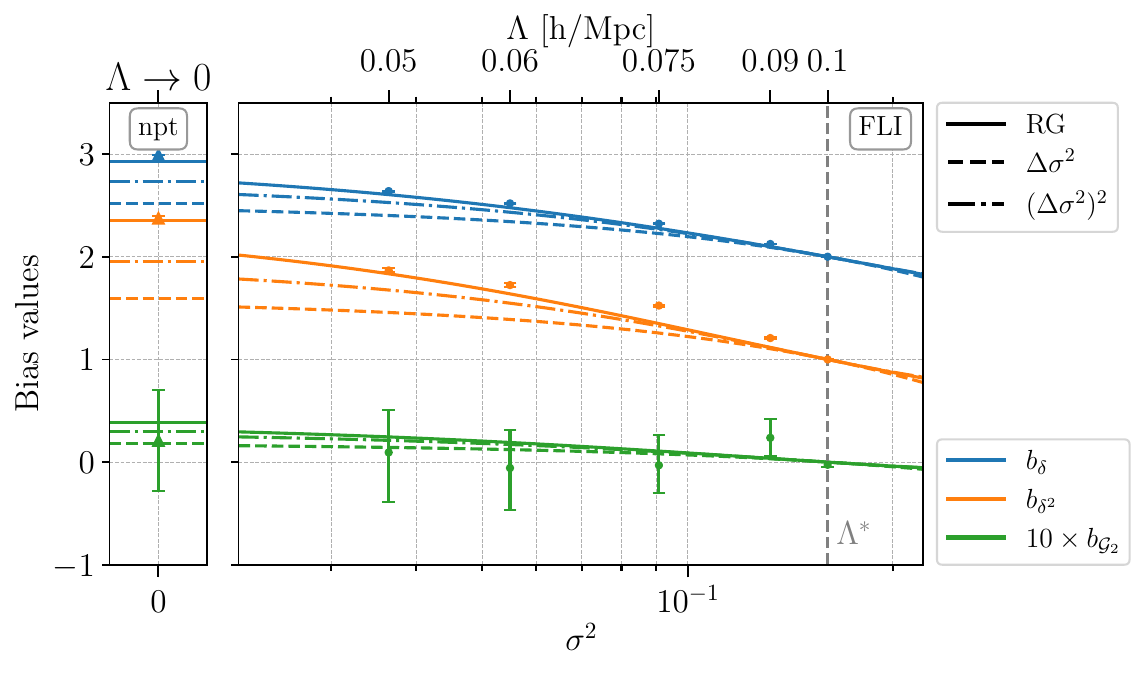}
    \caption{The cutoff dependence of $b_\delta,b_{\delta^2},$ and $b_{\mathcal{G}_2}$ measured by $n$-point functions (npt, left) and field-level inference (FLI, right), on mock tracer fields. The solid lines are the full prediction of one-loop RG equations \refeq{general_rg_equation}, dashed lines (dash-dotted lines) are solutions expanded up to $\Ds$ ($(\Ds)^2$) in \refeq{ODEsolution}. For the three cases we truncate the $s_{ab}^{\rm 1L}$ matrix at $s_{ab}^{[n\leq3]}$. The \texttt{LEFTfield} mocks are generated with the cutoff $\Lambda_*=0.1\ h/\mathrm{Mpc}$ on $\dlinL$ (vertical dashed line). Statistical error bars are computed using a set of 100 realizations for the field-level and $n$-point measurements, respectively (barely discernible for $b_\delta$ and $b_{\delta^2}$). }
    \label{fig:npt_FLI_4LPT}
\end{figure}

\reffig{npt_FLI_4LPT} shows the $n$-point function measurements at $\Lambda=0$ (left panel) and the field-level measurements at $\Lambda=0.05, 0.06, 0.075, 0.09, 0.1\ h/\mathrm{Mpc}$ (right panel). The results for $b_\delta$, $b_{\delta^2}$, and $b_{\mathcal{G}_2}$ are shown in different colors. As discussed in \refsec{trunc_rg}, we model gravitational evolution at 4LPT, which, as illustrated in \reffig{sourcing_diagram_1L}, captures the running of $b_\delta$ through order $(\Ds)^2$ and that of $b_{\delta^2}$ and $b_{\mathcal{G}_2}$ through order $\Ds$. We also present in \refapp{nLPT} results with lower-order LPT matter forward models.
The error bars for both field-level and $n$-point function bias measurements in \reffig{npt_FLI_4LPT} represent the uncertainty on the mean across 100 mock realizations.
We compare the measurements with RG predictions using the truncation described in \refsec{trunc_rg}, with the RG mixing matrix $s_{ab}^{[n\leq3]}$. 
We specify the initial conditions for the bias parameters at $\Lambda_*=0.1\ h/\mathrm{Mpc}$ (gray vertical line), where $\sigma^2_* \approx 0.16$  for this $\Lambda$CDM cosmology, such that nonlinear effects are relevant, but the perturbative description is still valid.
The line styles distinguish the RG predictions at different orders in $\Ds$, following \refeq{ODEsolution} (see also \reffig{fullRG}): dashed lines keep only the term of order $\Ds$; dash-dotted lines include terms up to order $(\Ds)^2$; and solid lines show the full RG solution within the truncation specified above.

In \reffig{npt_FLI_4LPT}, the RG predictions and mock measurements show the same overall trend, with the bias parameters rising when lowering $\Lambda$ and approaching a constant at low $\Lambda$.
The measured running of $b_\delta$ closely follows the full RG prediction not only close to $\Lambda_*$, but over the entire range of $\Lambda$, where $\Ds$ becomes comparable to $\sigma^2(\L_*)$. We also see in \reffig{npt_FLI_4LPT} that $b_{\mathcal{G}_2}$ agrees with the RG prediction, albeit within much larger relative error bars, due to the small numerical value of this coefficient. For $b_{\delta^2}$, the measurements generally agree with the prediction, although the deviations near $\Lambda_*$ exceed the estimated error bars.
We conjecture that the deviations are due to the fact that the RG sources higher-order and higher-derivative deterministic bias operators which are however missing in the forward model, and refer to \refapp{FLI_bias_estimation} for the analytical argument. 
In the limit $\Lambda \to 0$, the RG predictions  agree closely with the $n$-point function measurements for both $b_\delta$ and $b_{\delta^2}$, shown in the left panel of \reffig{npt_FLI_4LPT}. This agreement demonstrates that RG evolution can connect bias parameters inferred at finite cutoff in field-level analyses to those measured from $n$-point functions at $\Lambda=0$, allowing for consistent comparisons between the two approaches.

\subsection{The running of bias parameters on halos} \label{sec:res_halos}
%%%%%%%%%%%%%%%%%%%%%%%%%%%%%%%%%%%%%%%%%%%%%

We now turn to the halo catalogs described in \refsec{data}. Unlike the \texttt{LEFTfield} mocks, these catalogs have no known input bias parameters at a ground-truth cutoff $\Lambda_*$. We therefore fit the RG curves for $b_\delta$, $b_{\delta^2}$, and $b_{\mathcal{G}_2}$ to the measured bias parameters using the model
\be\label{eq:rg_fitting_model}
 \frac{d b_a}{d \sigma^2}=- {b_b} s_{ba}^{[n\leq3]} + B_a\,.
\ee
While for the mocks discussed in \refsec{res_mocks}, we could set higher-order bias parameters to zero at $\Lambda_*$, fourth- and higher-order bias parameters are generally nonzero for halos and can contribute to the running. We include their contributions to the running of the quadratic bias parameters by the constant source term $B_a=(0, B_{\delta^2}, B_{\mathcal{G}_2},0, 0, 0, 0)_a$ in \refeq{rg_fitting_model}, accounting for the terms omitted by truncating the sourcing matrix $s_{ba}^{[n\leq3]}$ at third order \cite{Rubira:2023vzw}. 
We initialize the RG flow of \refeq{rg_fitting_model} at $\Lambda=0$ given the boundary values of parameters,
\be
\{b^{\L=0}_{\delta},\  b^{\L=0}_{\delta^2},\  b^{\L=0}_{\mathcal{G}_2},\  b^{\L=0}_{\delta^3},\  b^{\L=0}_{\delta\mathcal{G}_2},\  B_{\delta^2},\  B_{\mathcal{G}_2}\}\,.
\ee
The first five of these are equivalent to the corresponding $\npt$-bias parameters, while the last two should be considered effective fitting parameters. 
The remaining third-order bias coefficients, $b_{\mathcal{G}_3}^{\L=0}$ and $b_{\Gamma_3}^{\L=0}$, cannot be constrained by our fits because their corresponding rows in \reftab{sab1L} vanish, so they do not source the running of the linear and quadratic bias parameters at the order considered. 
Moreover, at the order we work in, we do not expect to correctly describe the running of the third-order bias coefficients, so we do not show them.

We consider two scenarios for obtaining the best-fit values of the parameters listed above:  one that uses $n$-point function measurements to fix some of the boundary values at $\L=0$, and another that uses \emph{only} field-level measurements:
\begin{enumerate}
    \item \emph{$n$-pt anchored:} We fix the boundary values $b_{O^{\leq[2]}}^{\L=0}$ to the values from the $n$-point function measurements $b_{O^{\leq[2]}}^{n\text{-pt}}$, whose consistency with the bias relations is verified in \reffig{halo_bias_relation}. Using then the field-level measurements of $\hat{b}_{O^{[\leq2]},\mathrm{FLI}}^{\L}$ at different $\Lambda$, we fit $\{b_{\delta^3}^{\Lambda=0}, b_{\delta\mathcal{G}_2}^{\Lambda=0}, B_{\delta^2}, B_{\mathcal{G}_2}\}$.
    \item \emph{FLI only:} We additionally fit the three boundary values $b_{O^{\leq[2]}}^{\Lambda=0}$, together with the four parameters listed above, again using only the field-level measurements of $\hat{b}_{O^{[\leq2]},\mathrm{FLI}}^{\L}$ at different $\Lambda$.
\end{enumerate}
The first scenario is motivated by the fact that it provides the strongest consistency check between $n$-point and field-level bias measurements connected by the RG prediction. The second scenario shows the accuracy with which the $n$-point bias parameters can be predicted using the RG constrained by field-level measurements only.

\begin{figure}[t]
    \centering
    \includegraphics[width=0.49\linewidth]{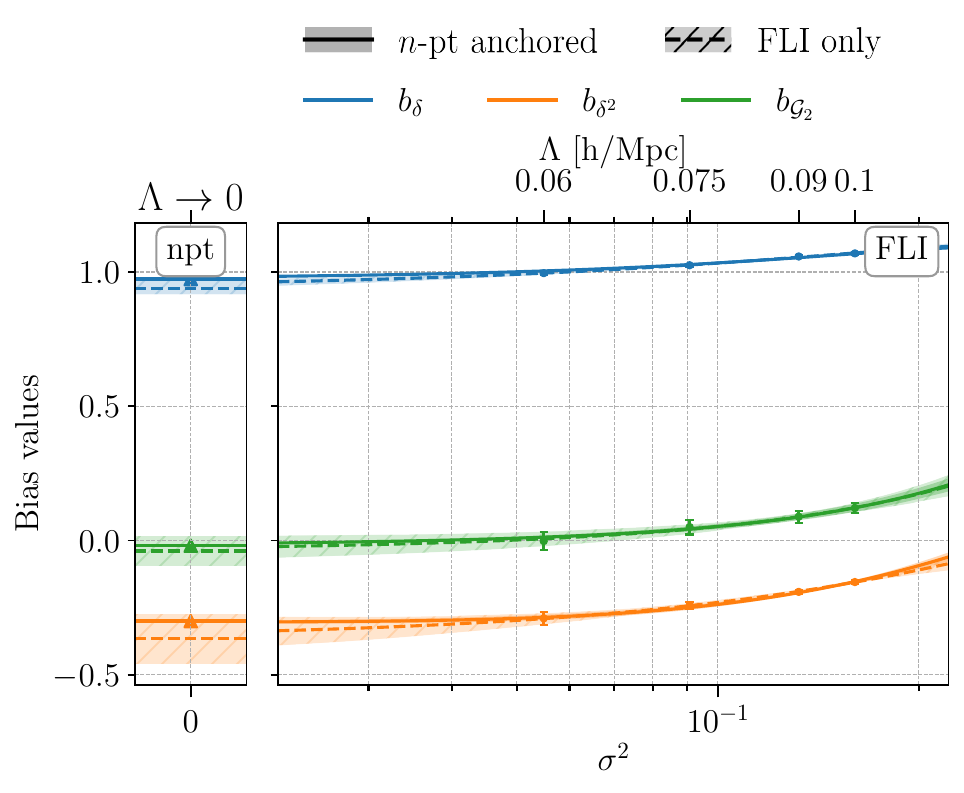}
    \includegraphics[width=0.49\linewidth]{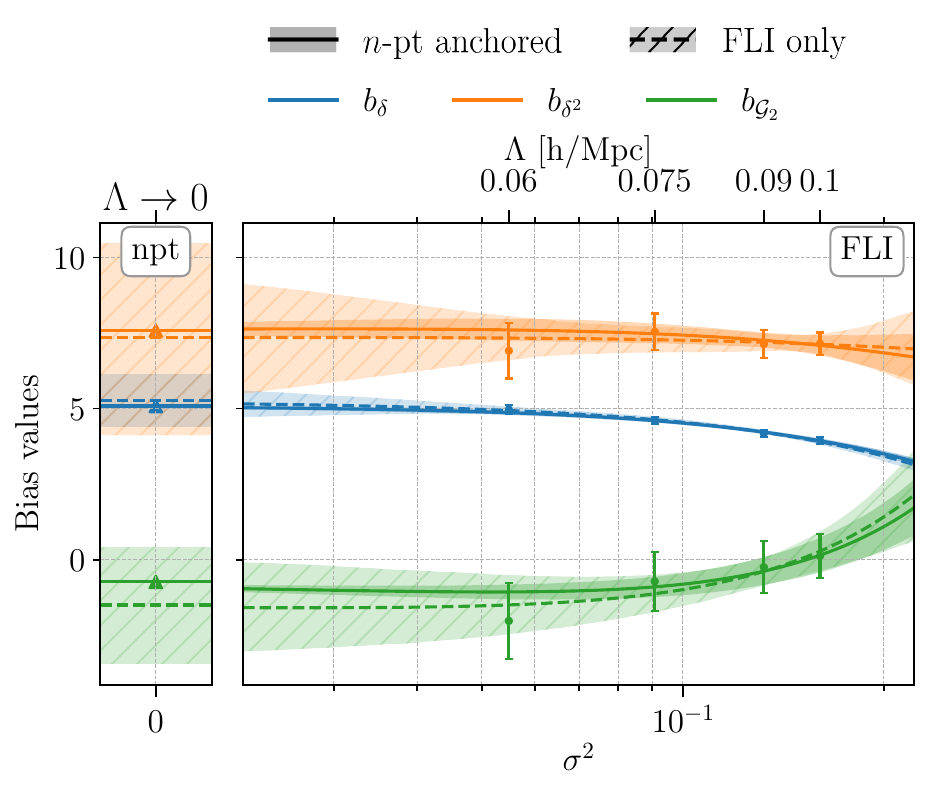}
    \caption{Running of the linear and quadratic bias parameters for dark matter halos. The left and right panels show the samples with $\log M=12.5\text{--}13.0$ at $z=0$ and $\log M=14.0\text{--}14.5$ at $z=1$, respectively. Curves show the RG fits with linear and quadratic bias parameters either fixed to the $n$-point function measurements (at $\Lambda=0$) or fitted using only the field-level measurements. Shaded bands indicate the estimated fit uncertainties (see main text for description).}
    \label{fig:halo_running}
\end{figure}

For each strategy, we maximize the likelihood
\be
\mathcal{L}({\bm{\theta}})\propto \exp\left[-\frac{1}{2}\sum_{\{\delta, \delta^2,\mathcal{G}_2\}} \sum_{\L_\mathrm{FLI}} \left( \frac{\hat{b}^\L_{O,\mathrm{FLI}}-b^\L_{O,\mathrm{RG}}(\bm{\theta})} {\sigma(\hat{b}^\L_{O,\mathrm{FLI}})} \right)^2\right],
\ee
where $\bm{\theta}$ contains the free parameters in each fitting setup, and the sum over $\L$ runs over the finite $\L$ values at which FLI measurements are available. Further, $\hat{b}^\L_{O,\mathrm{FLI}}$ and $\sigma(\hat{b}^\L_{O,\mathrm{FLI}})$ denote the field-level measurements and their uncertainties at cutoff $\Lambda$, and $b^\L_{O,\mathrm{RG}}(\bm{\theta})$ denotes the predictions obtained by solving \refeq{rg_fitting_model} with the corresponding boundary values and source terms. The resulting parameter uncertainties are estimated by the curvature around the best-fit point, following \refeq{ml_uncertainty}. 

We analyse the bias running for two halo samples drawn from the catalogs described in \refsec{data}, which have substantially different bias values: (1) mass bin $\log M=12.5\text{-}13.0$ at redshift $z=0$ (``low-mass sample'') and (2) mass bin $\log M=14.0\text{-}14.5$ at redshift $z=1$ (``high-mass sample''). These correspond to weakly and highly biased tracers, respectively. 
In order to include sufficient modes in the field-level analysis to constrain the RG fits, we increase $\kmax$ to $0.05\ h/\mathrm{Mpc}$ (from $0.025\ h/\rm Mpc$ employed for the mock catalogs) and perform measurements at $\Lambda=0.06, 0.075, 0.09, 0.1\ h/\mathrm{Mpc}$. 

\reffig{halo_running} shows the results, with each panel representing one halo sample, for the two fitting strategies, with solid lines for {\it $n$-pt anchored} and dashed lines for {\it FLI only}. The shaded bands show the fit uncertainties propagated through the RG running, which are determined as follows. We draw 2000 parameter sets $\bm{\theta}^{(k)}\sim\mathcal{N}(\hat{\bm{\theta}},C_\theta)$, where $\hat{\bm{\theta}}$ and $C_\theta$ are the best-fit parameters and their estimated covariance, evaluate the RG solution for each draw, and construct the $1\sigma$ interval at each $\Lambda$. 
The best-fit parameters and their uncertainties are also listed in \reftab{rg_fit_comparison}. Note that the $n$-point measurements are within the $1\sigma$ uncertainties estimated for the FLI-only bias inferences in all cases.

\begin{table}[h]
    \small
    \centering
    \caption{Comparison between the best-fit parameters $b_{O^{\leq[2]}}^{\L=0}$ for the fitting scenario `FLI only' and the $\npt$ best-fit bias parameters $b_{O^{\leq[2]}}^{\npt}$, applied to two different halo catalogs.
    Quoted uncertainties on FLI inferences correspond to the $1\sigma$ errors. 
}
    \label{tab:rg_fit_comparison}
    \begin{tabular}{lcccc}
        \toprule
        & \multicolumn{2}{c}{$\log M=12.5\text{-}13.0$ at $z=0$}
        & \multicolumn{2}{c}{$\log M=14.0\text{-}14.5$ at $z=1$} \\
        \cmidrule(lr){2-3}
        \cmidrule(lr){4-5}
        Parameter
        & $n$-pt best-fit
        & FLI only
        & $n$-pt best-fit
        & FLI only \\
        \midrule

        $b_\delta$
        & $0.974$ % \pm 0.005$
        & $0.940 \pm 0.027$ 
        & $5.087$ % \pm 0.126$
        & $5.265 \pm 0.916$ \\

        $b_{\delta^2}$
        & $-0.304 $ %\pm 0.014$
        & $-0.366 \pm 0.097$ 
        & $7.723$ % \pm 0.196$
        & $7.342 \pm 3.239$ \\

        $b_{\mathcal{G}_2}$
        & $-0.062$% \pm 0.041 $
        & $-0.039 \pm 0.054$ 
        & $-0.588$ % \pm 0.584$
        & $-1.50 \pm 1.90$ \\

        \bottomrule
    \end{tabular}
\end{table}

For the low-mass sample, the running with $\Lambda$ is relatively slow because $b_{\delta^2}$, the main coefficient driving the running, is small. Both the {\it $n$-pt anchored} and {\it FLI only} fits agree very well with the measured bias parameters. For the high-mass sample, the uncertainties in both the field-level and $n$-point function measurements are larger. Further, the larger bias coefficients lead to faster running with $\Lambda$, which results in wider uncertainty bands for the RG curves. Nevertheless, the fitted curves remain consistent with the measurements within their uncertainties, providing a good description of the running of all three bias parameters. 
Notice that the statistical error bars are larger compared to the results in \refsec{res_mocks}, since we analyze only a single realization.
The good agreement for both halo samples is reassuring, showing that the RG connection between finite-cutoff field-level bias parameters and $n$-point function measurements also holds for these fully nonlinear tracers.

\subsection{Comparison to previous field-level approaches}\label{sec:comparison_renormalization}

As mentioned, field-level bias constructions necessarily need a finite regulator in form of the grid employed. The approach described here so far makes explicit reference to the filtering
scale $\Lambda$ in the initial conditions. We now discuss how the grid resolution dependence was handled in previous field-level literature, broadly defined \cite{lazeyras_beyond_2018,Abidi_2018, schmidt_rigorous_2018, schmidt_unbiased_2020}.

Firstly, we study the connection with the \textit{transfer function} approach \cite{Schmittfull:2018yuk,Obuljen_2023,ivanov2024fullshapeanalysissimulationbasedpriors}, see \refapp{transfer_function} for a more detailed review. This approach builds the bias operators {out of the linear density field $\dlin(\vq)$, and shifts the operators from the Lagrangian position $\vq$ to the Eulerian position $\vx = \vq + \bm{s}_1$ using the Zel'dovich (first-order LPT) displacement $\bm{s}_1$. The operator set is extended with a few extra operators by matching up to a specific perturbative order. Explicitly, the Fourier-space expression for an operator $O$ given linear density field and displacement filtered on the scale $\L$ is}
\be\label{eq:shifted_operator}
\tO{O}_\L(\vk)=\int_{\vq} O[\dlinL](\vq) e^{-i\vk\cdot[\vq+\bm{s}_{1,\L}(\vq)]},
\ee 
where mean subtraction is done on the operators before shifting. 
Refs.~\cite{Schmittfull:2018yuk,Obuljen_2023} match the coefficients in front of the shifted operators with the standard bias expansion by expanding the exponential in \refeq{shifted_operator} and matching its kernel
\be\label{eq:shifted_operator_kernel}
K_{\tO{O}}^{(n)}=\frac{1}{(n-m)!}\ \mathrm{Sym}\left[K_{O}^{(m)}(\vp_1,...,\vp_m)\prod_{a=m+1}^n\frac{\vk\cdot\vp_a}{p_a^2}\right],\quad n\geq m,
\ee 
to the general bias expansion \refeq{operator_kernel}, which in the existing literature has been done up to third order in $\dlin$ (Sec. VI in \cite{Schmittfull:2018yuk}). 
The aim is then to remove the grid-scale dependence by enforcing the conditions in
\refeq{renormalization_conditions}. For this, Refs. \cite{Schmittfull:2018yuk,ivanov2024fullshapeanalysissimulationbasedpriors} promote the coefficients of the shifted operators to $k$-dependent transfer functions $\beta_O(k)$.
These can be computed approximately using perturbation theory, or using simulations,
after which, ideally, the quantities $\lim_{k\to 0} \beta_O(k)$ match the $n$-point-renormalized $b_O$ (in fact, a rotated bias basis is used, see \refapp{transfer_function} for the technical details).

This approach has several drawbacks. First, \refeq{shifted_operator} is not complete, as it does not capture the displacement $\bm{s}$ beyond linear order. This means that additional terms need to be added to the set of shifted operators. While simple at third order, where only a single term needs to be added, this becomes more complicated at higher orders, and there is no automatic guarantee of completeness of the bias expansion.

Second, the transfer function computation in perturbation theory is significantly more tedious than the shell integrals needed for the RG approach. Indeed, the full 1-loop computation of the transfer functions for second-order bias terms has not been done. Moreover, since the transfer functions are only computed for the bias terms in a predefined list, they do not yield independent confirmation of the completeness of the bias expansion.
It is worth noting that, in any case, a robust connection of the transfer functions to $n$-point-renormalized bias coefficients is only possible if the initial conditions cutoff is safely in the perturbative regime, which is not the case for the analysis in \cite{ivanov2024fullshapeanalysissimulationbasedpriors}, for example; indeed, the authors acknowledge a residual cutoff dependence of the inferred $n$-point-renormalized bias coefficients.

The simulation-based measurement of the transfer functions on the other hand is costly and makes the incorporation of the cosmology-dependence of the cutoff removal difficult, in addition to the fact that it does not offer analytical insights and completeness checks.

Another previous line of work \cite{lazeyras_assembly_2021} constructed the operators in a similar way as \refsec{forward_model}, but applied a renormalization to the quadratic operators by subtracting $\delta(\vk)$ times a $k$-dependent function computed at 1-loop order \cite{schmidt_rigorous_2018}. 
This operation merely removes the leading cutoff-dependent contribution, and there is residual dependence on the cutoff scale $\Lambda$, see the discussion in Sec. 4.1. in \cite{lazeyras_assembly_2021}.  To absorb the residual $\L$ dependence, they fitted $b_O(\L)$ with a function $b_O^{\npt} + \L^2 b_O^{(2)}$, leaving the effective $b_O^{(2)}$ free, to obtain estimates for the $\npt$ bias parameters in the $\L\to 0$ limit. 
Finally, Ref.~\cite{lazeyras_beyond_2018} only uses the Gaussian counterterm subtraction.

%%%%%%%%%%%%%%%%%%%%%%%%%%%%%%%%%%%%%%%%%%%%%%%%%%%%%%%%%%%%%%%%%%%%%%%%%%%
%%%%%%%%%%%%%%%%%%%%%%%%%%%%%%%%%%%%%%%%%%%%%%%%%%%%%%%%%%%%%%%%%%%%%%%%%%%
\section{Conclusion}\label{sec:conclusion}

In this work, we connect the renormalized bias parameters measured from $n$-point functions to the finite-cutoff, $\Lambda$-dependent bias parameters inferred at the field level using the renormalization group (RG) of galaxy bias \cite{Rubira:2023vzw,rubira_renormalization_2024,Bakx:2025cvu}. We present results for the running of the first- and second-order bias parameters $b_\delta$, $b_{\delta^2}$, and $b_{\mathcal{G}_2}$. We consider a systematic truncation of the RG running at a fixed order in $\Ds$ and identify the corresponding truncation in LPT order required to achieve this accuracy (see \refsec{trunc_rg} and \reffig{sourcing_diagram_1L}). We use a 4LPT matter forward model with a third-order bias expansion and a Gaussian likelihood in which we marginalize over the third-order bias parameters.

We test the running of the bias parameters defined at a lattice cutoff in two setups, first using controlled perturbative galaxy mocks generated with \texttt{LEFTfield} and then using halo catalogs from $N$-body simulations. For the mocks, we find good agreement between the theoretical RG predictions and both the $n$-point function and field-level measurements, as shown in \reffig{npt_FLI_4LPT}.
The tests using halo catalogs represent a more realistic scenario, in which the initial values of the bias parameters at a fixed cutoff $\Lambda$ are not known. We fit the RG curves using \refeq{rg_fitting_model} and again find good agreement with the measurements for both low- and high-mass samples, as shown in \reffig{halo_running}. The low-mass samples have smaller bias parameters and exhibit slower running with $\Lambda$, being less sensitive to higher-order terms in $\Ds$ and correspondingly presenting smaller uncertainties in the RG prediction. The high-mass sample has larger bias parameters and greater sensitivity to higher-order terms, which translates into larger uncertainties in the predicted running. Again, the field-level measurements agree well with the predicted running within these error bars.

We also provide the connection  with the transfer-function approach previously used to approximately remove the cutoff-dependence in field-level bias constructions (\refsec{comparison_renormalization}). 
A key advantage of the RG calculation presented here is that it is more systematic and allows for controlled predictions of the running, including estimated uncertainties from higher order terms.

In conclusion, our results fully describe the cutoff dependence of the bias running, both for controlled mock samples and for halo catalogs. Moreover, by estimating the effect of higher-order bias contributions neglected in the RG calculation, we can provide a reliable systematic error bar on the inferred running (\reffig{halo_running}).
This work shows that using the galaxy-bias RG is a natural way to connect finite-cutoff field-level measurements to renormalized bias parameters from $n$-point functions in the limit $\Lambda \to 0$, providing a framework for comparing bias measurements across different analysis schemes and for constructing consistent bias priors for galaxy-survey analyses. 
We emphasize that an accurate prediction of the running from finite cutoff to $\L\to 0$ is crucial in order to reliably use field-level measurements to inform $\npt$-renormalized bias parameters.

While we have focused on the \texttt{LEFTfield} Eulerian bias forward model here (i.e. bias operators constructed out of $\delta_{\rm n\text{-}LPT}$), the same techniques can be applied to any complete field-level bias operator construction, e.g. a purely Lagrangian construction \cite{schmidt_n-th_2021}.
In the future, we plan on extending this comparison beyond second-order bias parameters, analyzing the running of stochastic parameters and exploring further applications of bias running in galaxy-survey analyses.

\acknowledgments
Hsiang-Ming Huang thanks Ivana Nikolac, Beatriz Tucci and Julia Stadler for useful discussions on \texttt{LEFTfield}.
All numerical computations were performed on the HPC cluster \href{https://docs.mpcdf.mpg.de/doc/computing/clusters/systems/Astrophysics/MPA-FREYA.html}{\texttt{FREYA}}, maintained by the \href{https://www.mpcdf.mpg.de/}{Max Planck Computing $\&$ Data Facility}.

\textit{AI usage statement}: ChatGPT (GPT-5.6) was used to optimize the original code written by a human, to improve individual formulations, and to perform grammar checks in the paper. The authors have verified all results and take full responsibility for all results presented herein.

\appendix

%%%%%%%%%%%%%%%%%%%%%%%%%%%%%%%%%%%%%%%%%%%%%%%%%%%%%%%%%%%%%%%%%%%%%%%%%%%%
\section{Relation between LPT and SPT kernels}\label{app:LPT_order}

In this appendix we review the relation between the LPT and SPT descriptions of the perturbative matter density field. 
We decompose the displacement into longitudinal and transverse components,
\ba
\bm{s}_L(\vk,\tau) &= -i \sum_n D^n(\tau) \bm{L}^{(n)}(\vk),\\
\bm{s}_T(\vk,\tau) &= -i \sum_n D^n(\tau) \bm{T}^{(n)}(\vk),
\ea
where the $n$-th order contributions can be written in terms of the symmetrized longitudinal and transverse kernels $S_L^{(n)}$ and $\bm{S}_T^{(n)}$ as
\bea
\bm{L}^{(n)}(\vk)=\int_{\vp_1,\dots,\vp_n}\diracpi(\vp_{1...n}-\vk)\frac{\vk}{k^2}{S}_L^{(n)}(\vp_1,\dots,\vp_n)\dlin(\vp_1)\cdots\dlin(\vp_n),\label{eq:LPT_kernel_def_L}\\
\bm{T}^{(n)}(\vk)=\int_{\vp_1,\dots,\vp_n}\diracpi(\vp_{1...n}-\vk)\frac{1}{k^2}\vk\times\bm{S}_T^{(n)}(\vp_1,\dots,\vp_n)\dlin(\vp_1)\cdots\dlin(\vp_n).\label{eq:LPT_kernel_def_T}
\eea
At first order, $S_L^{(1)}=1$ and $\bm{S}_T^{(1)}=0$. The transverse part first appears at third order, i.e. $\bm{S}_T^{(2)}=0$.

Expanding the exponential $e^{i\vk\cdot\bm{s}}$ in \refeq{density_assignment_fourier} in powers of the displacement and matching to the SPT kernels order by order, we find, up to $n=3$ \cite{Rampf_2012a,Rampf_2012b},
\ba \label{eq:StoF}
F_1(\vp_1)=\, &\vk\cdot\bm{S}^{(1)}_{L\oplus T}(\vp_1),\\ 
F_2(\vp_1,\vp_2)=\, &\vk\cdot \bm{S}^{(2)}_{L\oplus T}(\vp_1,\vp_2)+\frac{1}{2}\vk\cdot\bm{S}_{L\oplus T}^{(1)}(\vp_1)\vk\cdot\bm{S}_{L\oplus T}^{(1)}(\vp_2), \nonumber\\ 
F_3(\vp_1,\vp_2,\vp_3)= \, &\vk\cdot\bm{S}_{L\oplus T}^{(3)}(\vp_1,\vp_2,\vp_3) +\frac{1}{3}\left[\vk\cdot\bm{S}_{L\oplus T}^{(1)}(\vp_1)\vk\cdot\bm{S}_{L\oplus T}^{(2)}(\vp_2,\vp_3)+\mathrm{two\ perms.}\right], \nonumber \nonumber \\ 
&+\frac{1}{6}\vk\cdot\bm{S}_{L\oplus T}^{(1)}(\vp_1)\vk\cdot\bm{S}_{L\oplus T}^{(1)}(\vp_2)\vk\cdot\bm{S}_{L\oplus T}^{(1)}(\vp_3), \nonumber
\ea 
with $\vk=\vp_{1...n}$ and where, as shorthand,
\be
\bm{S}^{(m)}_{L\oplus T}(\vp_1,\dots,\vp_m)\equiv\frac{\vp_{1...m}}{p_{1...m}^2}\,S^{(m)}_{L}(\vp_1,\dots,\vp_m)+\frac{\vp_{1...m}\times\bm{S}^{(m)}_{T}(\vp_1,\dots,\vp_m)}{p_{1...m}^2}\,.
\ee

%%%%%%%%%%%%%%%%%%%%%%%%%%%%%%%%%%%%%%%%%%%%%%%%%%%%%%%%%%%%%%%%%%%%%%%%%%%%
\section{The effect of the incomplete forward model on the field-level inference}\label{app:FLI_bias_estimation}

In this appendix, we discuss two possible sources of systematic errors in the fixed-phase field-level bias measurements.
The first issue is whether the Gaussian likelihood employed in the measurement biases the parameter estimation, given that the actual noise is in general non-Gaussian and density-dependent. The second issue is bias operators not included in the deterministic forward model employed in the measurement.  We discuss them together in the following.

The tracer density field $\hdg$ on which the inference is performed can generally be decomposed as follows:
\be\label{eq:true_dg}
\hdg=\sum_{O_\mathrm{mod}} b_{O,\mathrm{true}} O + \delta_{g,\mathrm{mis}} + \eta, 
\ee 
where $O_\mathrm{mod}$ are the set of deterministic bias operators used in the forward model, $\delta_{g,\mathrm{mis}}$ is the part missing from the model, e.g. higher-order $O^{[4]}$ or next-to-leading gradient order $\lapl(\delta^2)$, and $\eta$ is the general noise field. While the likelihood used in the inference
assumes a Gaussian noise field $\epsilon$, in general the noise field is written as $\eta = \epsilon + b_{\epsilon,\delta} \epsilon\delta + b_{\epsilon^2}\epsilon^2 + \cdots$ \cite{rubira2025nongaussiangalaxystochasticitynoisefield}.
Note that the general noise field still obeys $\langle \eta(\vk)O(\vk')\rangle_{\eta,\dlin} = 0$ for all bias operators $O$.
On the other hand, the tracer density field predicted by the model is given as (\refsec{forward_model})
\be 
\delta_g = \sum_{O_\mathrm{mod}} b_O O +\epsilon\,.
\ee 
By taking the maximum likelihood (ML) condition for $b_{O_\mathrm{mod}}$ and $\sigma_\epsilon^2$ from the Gaussian likelihood \refeq{likelihood}, we obtain the estimator for the best-fit parameters \cite{schmidt_rigorous_2018},
\ba 
\hat{{b}}_{O_a,\mathrm{ML}} & =  \sum_{O_b}({M}^{-1})_{ab} H_b \,,\label{eq:b_ML}\\ 
\hat{\sigma^2_{\epsilon}}_\mathrm{ML} & = 
\frac{\sum_{\vk} r^*(\vk)r(\vk)}{N} \label{eq:sigma2_ML},
\ea 
where $M_{ab}=\sum_{\vk}  O_a^{*}(\vk)O_b(\vk)$,  $H_a= \sum_{\vk} O_a^{*}(\vk)\hdg(\vk)$,\footnote{A constant $\sigma_\epsilon$ has been canceled in $H$ and $M$.} $N$ is the number of modes and we introduce the residual
\ba\label{eq:residual_field}
r(\vk) & =\hdg(\vk)- \sum_{O_\mathrm{mod}}b_{O,\mathrm{ML}} O(\vk) \\ 
& = \sum_{\vk'} R_{\vk,\vk'} \left[\delta_{g,\mathrm{mis}}(\vk')+\eta(\vk')\right] \vs
\mbox{where}\quad
R_{\vk,\vk'}& \equiv {\dirac}^{\vk,\vk'}-\sum_{O_a,O_b}O_a(\vk)({M}^{-1})_{ab}{O}_b^*(\vk').
\ea 
In the second line, we used \refeq{b_ML} and \refeq{true_dg}. Notice that $R_{\vk,\vk'}$ is a projection operator satisfying $\bm{R}^T=\bm{R}$ and $\bm{RR}=\bm{R}$. Using the zero conditional mean of $\eta$ on a fixed $\dlinL$, i.e. $\langle\eta|\dlinL\rangle_{\eta} = 0$ from the definition of $\eta$, we can obtain the expectation values over the ensemble of noise and initial condition $\dlinL$:
\ba  
\mathbb{E}_{\eta,\dlinL}[\hat{b}_{O_a,{\mathrm{ML}}}] - {b}_{O_a,\mathrm{true}} & = \mathbb{E}_{\dlinL}\left[\sum_{O_b}({M}^{-1})_{ab}\sum_{\vk}O_{b}^*(\vk) \delta_{g,\mathrm{mis}}(\vk)\right] \\
\mathbb{E}_{\eta,\dlinL}[\hat{\sigma^2_{\epsilon}}_\mathrm{ML}] & = \frac{1}{N}\mathbb{E}_{\dlinL}\left[\sum_{\vk,\vk'}R_{\vk,\vk'} C_{\mathrm{true},\vk,\vk'}+\sum_{\vk,\vk'}\delta_{g,\mathrm{mis}}^*(\vk)R_{\vk,\vk'}\delta_{g,\mathrm{mis}}(\vk')\right],\label{eq:exp_sigma2}
\ea 
The covariance matrix $C^{\mathrm{true}}_{\vk,\vk'}\equiv\langle\eta^*(\vk)\eta(\vk')\rangle$ can be rewritten at leading order in terms of the Gaussian stochastic parameters $P_{\epsilon,\unit}\diracpi(\vk-\vk')+\sum_{O}P_{\epsilon,O}O(\vk-\vk')$, and it gives $P_{\epsilon,\unit}$ merely under the trace computation
$\sum_{\vk}R_{\vk,\vk}\approx N$. The second contribution in \refeq{exp_sigma2} is positive definite since $\bm{R}=\bm{R}^T\bm{R}$.
These results show that missing terms in the bias expansion in general bias the estimated bias parameters, if there is overlap between the operators $O$ included and $\delta_{g,\rm mis}$. Further, the power spectrum of $\delta_{g,\rm mis}$ contributes to the estimated noise variance. Importantly, the Gaussian noise assumption does not bias the maximum-likelihood values of the bias parameters in the ensemble mean.

%%%%%%%%%%%%%%%%%%%%%%%%%%%%%%%%%%%%%%%%%%%%%%%%%%%%%%%%%%%%%%%%%%%%%%%%%%%%

\section{Convergence tests of the forward model}

In this appendix, we discuss convergence tests to find the optimal setup for creating and measuring the RG flows. The setup can be divided into two parts: (1) forward model and (2) field-level inference likelihood. For the former, we study the effect of $n$LPT matter forward model in \refapp{nLPT} and the grid resolution of $\delta_\mathrm{fwd}$ in \refapp{resolution_effect}. As for the likelihood, we study whether the marginalization and the truncation of the bias expansion affect the inference.

In the following, the common setup used in the comparison is: 4LPT matter forward model, with $N_\mathrm{G,Eul}=N_\mathrm{G,Ny}(4\Lambda)$ Eulerian resolution grid, third-order Eulerian bias expansion, and analytical marginalization over third-order bias parameters at the likelihood level.

\subsection{LPT order comparison}\label{app:nLPT}
\begin{figure}
    \centering
    \includegraphics[width=0.7\linewidth]{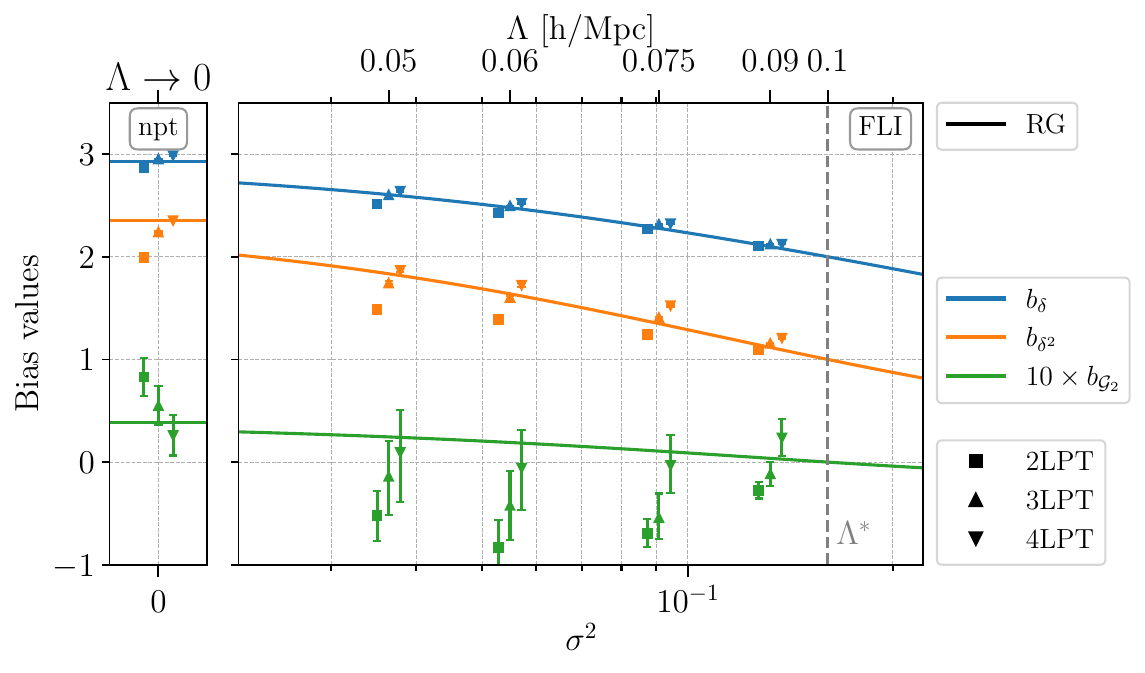}
    \caption{ Running of the bias parameters with different $n$LPT matter forward models: 2LPT (squares), 3LPT (upward triangles) and 4LPT (downward triangles). 2LPT and 4LPT points are slightly shifted horizontally for the visualization.}
    \label{fig:nLPT_summary}
\end{figure}

The perturbative order of the matter forward model affects the running of the bias parameters, as discussed in \refsec{trunc_rg}. We compute the matter field in $n$LPT for $n = 2, 3, 4$. Both mocks and samples are created at same perturbative order for each $n$LPT. We always keep $N_\mathrm{G,fwd}=N_\mathrm{G,Eul} = N_\mathrm{G,Ny}(n_\mathrm{LPT} \Lambda)$ for the evolution part of \refsec{forward_model}.

As shown in \reffig{sourcing_diagram_1L}, each higher LPT order adds nonlinear contributions to the bias running. 4LPT is the lowest order that captures the running of $b_\delta$ at $\mathcal{O}((\Ds)^2)$ and of the second-order bias parameters at $\mathcal{O}(\Ds)$.
\reffig{nLPT_summary} shows the results, indicating that the perturbative order of the matter forward model affects the bias running. 

%%%%%%%%%%%%%%%%%%%%%%%%%%%%%%%%%%%%%%%%%%%%%%%%%%%%%%%%%%%%%%%%%%%%%%%%%%%%
\subsection{Resolution of Eulerian grid}\label{app:resolution_effect}
We examine how the resolution of the Eulerian matter grid affects the measured running of the bias parameters. Although the likelihood only includes large-scale modes, the Eulerian bias operators are constructed from products of the evolved matter field. Short-scale matter modes can therefore contribute to large-scale bias operators through mode coupling and removing these modes before constructing the operators can modify the measured running. Similar effects in field-level inference have been studied in \cite{schmidt_unbiased_2020,stadler_fast_2025}, which we follow closely here. 

The NUFFT assignment computes the modes below the Nyquist frequency of the grid exactly and discards those above it, so its only effect is a filtering of $\delta_\mathrm{fwd}$. This truncation matters only through the nonlinear bias operators, in which modes of $\delta_\mathrm{fwd}$ above the Nyquist frequency could combine into the modes $|\vk|<k_\mathrm{max}$ entering the likelihood. Since the $n$-th order matter field is a convolution of $n$ linear fields with support $|\vp_i|<\Lambda$,
\be
\delta^{(n)}(\vk)=\int_{\vp_1,\ldots,\vp_n}\diracpi(\vk-\vp_{1\ldots n})\,F_n(\vp_1,\ldots,\vp_n)\,\dlinL(\vp_1)\cdots\dlinL(\vp_n)=0\quad\text{for}\quad k>n\Lambda\,,
\ee
where for $n\leq4$ the 4LPT kernels coincide with the SPT kernels $F_n$ (see \refapp{LPT_order}). We adopt the choice $N_\mathrm{G,Eul}=N_\mathrm{G,Ny}(4\Lambda)$, which resolves $\delta^{(n\leq4)}$, and every operator built from it, exactly for arbitrary momentum configurations. This is sufficient for \refeq{bias_rg_relation_matter_kernel}, which involves kernels up to $F_4$.

Considering a 4LPT model as the main text, the matter density field obtained after the density assignment on a grid with the size $N_{G,\mathrm{Eul}}=N_{G,\mathrm{Ny}}(4\Lambda_\mathrm{in})$ is 
\ba\label{eq:delta_fwd_4lpt}
\delta_\mathrm{fwd,4LPT}(\vk)&=W(k;4\Lambda_\mathrm{in})\int d^3q e^{-i\vk\cdot\vq}[e^{-i\vk\cdot\bm{s}^{(n\leq4)}(\vq,\lambda)}-1] \nonumber\\
&=W(k;4\Lambda_\mathrm{in})\sum_{n=1}^{\infty}\int_{\vp_1,...,\vp_n}K^{\mathrm{4LPT}, (n)}(\vp_1,...,\vp_n)\dlinLin(\vp_1)...\dlinLin(\vp_n)
\ea
where $W(k;4\Lambda_\mathrm{in})$ is a sharp-$k$ filter, $\v{s}^{(n\leq4)}$ is the 4LPT displacement field and its kernel $K^{\mathrm{4LPT}, (n)}$. $K^{\mathrm{4LPT}, (n)}$ are the same as the normal SPT kernels up to 4th-order as discussed in \refapp{LPT_order}.
The operator $\delta^2$ constructed with \refeq{delta_fwd_4lpt} is 
\ba
\delta^2(\vk)=&\int_{\vp}\delta_\mathrm{fwd,4LPT}(\vp)\delta_\mathrm{fwd,4LPT}(\vk-\vp) \nonumber\\
=&\int_{\vp}W(p;4\Lambda_\mathrm{in})W(|\vk-\vp|;4\Lambda_\mathrm{in})\sum_{n=1}^{\infty}\sum_{m=1}^{\infty}\int_{\vp_1,...,\vp_n,\vq_1,...,\vq_m} \diracc (\vp-\vp_{1...n})\diracc (\vk-\vp-\vq_{1...n}) \nonumber\\
&K^{\mathrm{4LPT}, (n)}(\vp_1,...,\vp_n)K^{\mathrm{4LPT}, (m)}(\vq_1,...,\vq_m)\dlinLin(\vp_1)...\dlinLin(\vp_n)\dlinLin(\vq_1)...\dlinLin(\vq_m)
\ea
Comparing the above equation with the shell correlator in \refeq{bias_rg_relation_matter_kernel}, we notice that the filter truncates the kernels starting from $5$th order, which is acceptable for the study in the main text.

\begin{figure}
    \centering
    \includegraphics[width=0.8\linewidth]{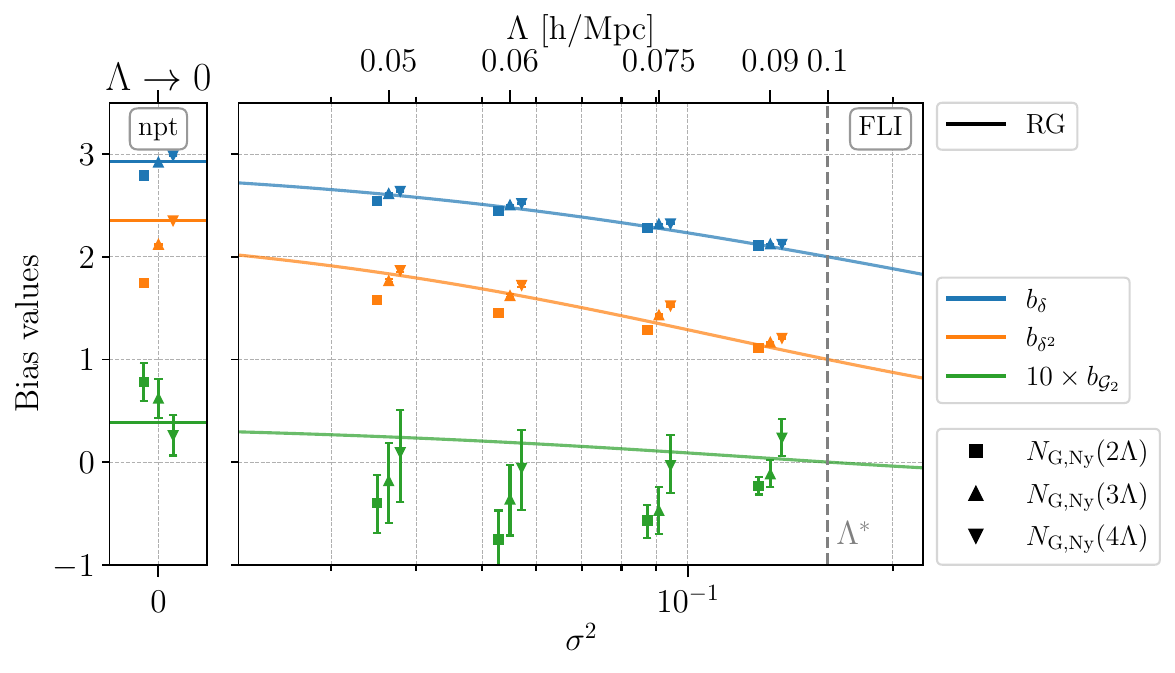}
    \caption{The resolution effect from the matter grid size $N_{G,\mathrm{Eul}}=N_\mathrm{G,Ny}(n\Lambda)$ on the magnitude of running. $n$ determines the size of the grid. $N_{\mathrm{G,Eul}}=N_\mathrm{G,Ny}(2\Lambda)$ and $N_\mathrm{G,Ny}(4\Lambda)$ points are slightly shifted horizontally for the visualization.}
    \label{fig:2LPT_NGeul_2NGin_sOOn2}
\end{figure}

Here, we briefly show the resolution effect with a 4LPT matter forward model with $N_{G,\mathrm{Eul}}=N_{G,\mathrm{Ny}}(n\Lambda_\mathrm{in})$ for $n=2,3$ and 4, see \reffig{2LPT_NGeul_2NGin_sOOn2}.
For $n=2$ and 3, the $F_3$ kernel is filtered following the same logic above and the sourcing from $\delta^2$ to $O^{[3]}$ in the second line of \refeq{bias_rg_relation_matter_kernel} are more suppressed. Therefore they are closer to the RG predictions with the mixing matrix $s^{[n\leq2]}$.

This condition is conservative since, for instance, the sourcing of $O^{[3]}$ by $\delta^2$ comes from its fifth-order part, $\left[\delta^2\right]^{(5)}=2\delta^{(1)}\delta^{(4)}+2\delta^{(2)}\delta^{(3)}$, where each term is a convolution $\big[\delta^{(a)}\delta^{(b)}\big](\vk)=\int_{\vp}\delta^{(a)}(\vp)\,\delta^{(b)}(\vk-\vp)$. Since $\delta^{(a)}(\vp)$ vanishes for $|\vp|>a\Lambda$, the factor $\delta^{(b)}$ is only probed at
\begin{equation}
|\vk-\vp|\leq|\vp|+k\leq a\Lambda+k\,.
\end{equation}
Hence, $\delta^{(4)}$ is only needed up to $\Lambda+k$ and $\delta^{(3)}$ up to $2\Lambda+k$, although their support extends to $4\Lambda$ and $3\Lambda$, respectively. Modes beyond these bounds cannot couple to produce modes with $|\vk|<k_\mathrm{max}$. The same holds for $\mathcal{G}_2$ and the cubic operators, so the RG relation only requires $\delta_\mathrm{fwd}$ up to $2\Lambda+k_\mathrm{max}<3\Lambda$. Once $N_\mathrm{G,Eul}$ is fixed, the bias operators are constructed on a grid of size $N_\mathrm{G,final}=3N_\mathrm{G,Eul}$. Since $\delta_\mathrm{fwd}$ is supported in $|k_i|\leq k_\mathrm{Ny}(N_\mathrm{G,Eul})$ along each axis, the cubic operators are supported in $|k_i|\leq3k_\mathrm{Ny}(N_\mathrm{G,Eul})$, which is fully resolved in this grid, so that no aliasing occurs in their construction. 

We compare in \reffig{2LPT_NGeul_2NGin_sOOn2} the running obtained with $N_\mathrm{G,Eul}=N_\mathrm{G,Ny}(n\Lambda)$ for $n=2,3,4$, keeping $N_\mathrm{G,final}=3N_\mathrm{G,Eul}$. For $n=3$, all contributions to \refeq{bias_rg_relation_matter_kernel} are retained, and the result is very close to the fiducial case $n=4$, and the small offset could be attributed to the fact that the NUFFT scheme does not behave exactly as a sharp-$k$ filtering. Note that the $n=4$ case closely reproduces the $n$-point function when $\Lambda \to 0$. For $n=2$, instead, $\delta^{(3)}$ is truncated for $2\Lambda<|\vp|<2\Lambda+k$, so the $\delta^{(2)}\delta^{(3)}$ contribution is already incomplete at leading order, consistent with the larger deviation in this case.

%%%%%%%%%%%%%%%%%%%%%%%%%%%%%%%%%%%%%%%%%%%%%%%%%%%%%%%%%%%%%%%%%%%%%%%%%%%%
\subsection{Prior volume in higher-order bias parameters}\label{app:marginalization}

\begin{figure}
    \centering
    \includegraphics[width=0.48\linewidth]{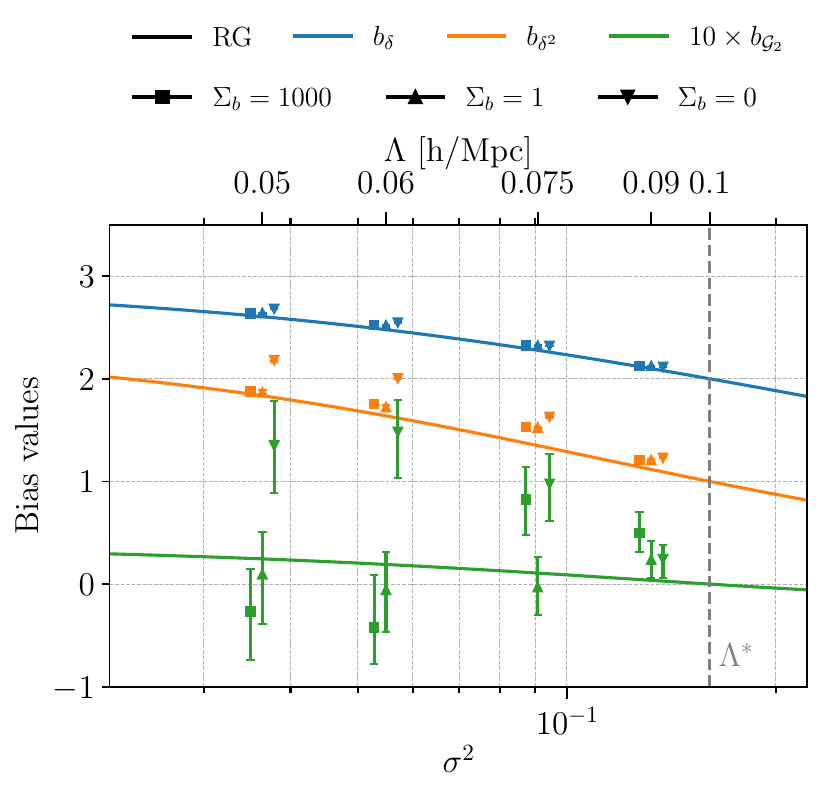}
    \includegraphics[width=0.48\linewidth]{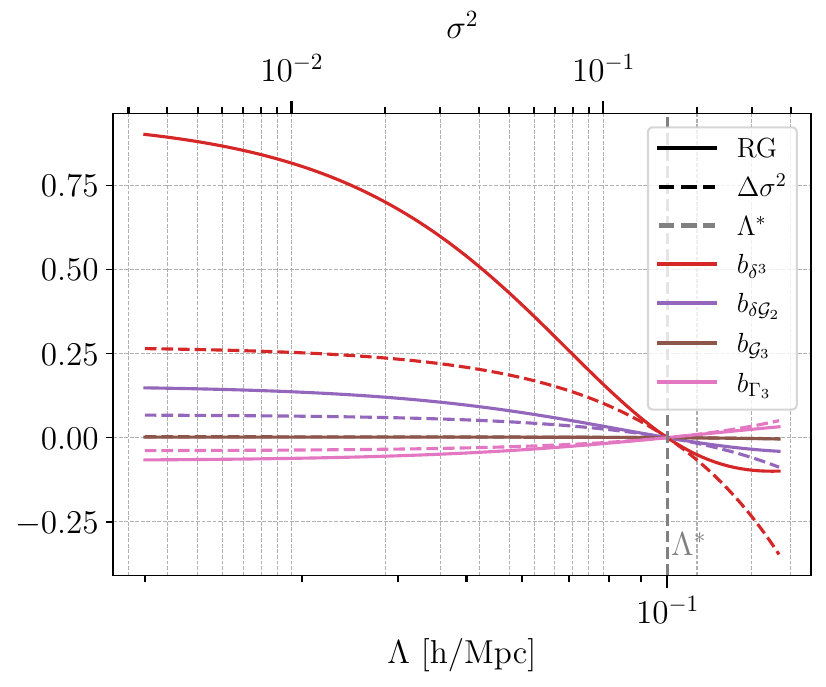}
    \caption{Left: effect of different choices of prior on the third-order bias parameters on the first- and second-order parameters. Markers show the bias parameters inferred at several cutoffs $\Lambda < \Lambda_*$ with the three likelihood cases described in the text: a third-order likelihood with a prior $\Sigma_b=1000$ and $\Sigma_b=1$ over third-order parameters and a second-order likelihood with no third-order coefficients (which effectively corresponds to a singular $\Sigma_b=0$ prior). Right: RG running of the third-order bias parameters $\{b_{\delta^3}, b_{\delta \mathcal{G}_2}, b_{\mathcal{G}_3}, b_{\Gamma_3}\}$ for the same initial conditions as the mocks in the left panel, $(b^{\Lambda_*}_\delta, b^{\Lambda_*}_{\delta^2}, b^{\Lambda_*}_{\mathcal{G}_2},\sigma_\epsilon^{\Lambda_*})=(2,1,0,0.2)$ at $\Lambda_*=0.1\,h/\mathrm{Mpc}$, with all third-order biases set to zero at $\Lambda_*$. }
    \label{fig:marg_effect}
\end{figure}

As discussed in \refsec{field_measure}, we expect the inference to become biased when the RG running generates non-negligible third- and higher-order bias parameters. 
As argued in \refsec{trunc_rg}, given the RG truncation, the solution \refeq{ODEsolution} predicts the third-order parameters only at leading order in $\Ds$.
To quantify this effect, we compare the bias parameters inferred in three cases:
(1) the third-order bias expansion, in which the third-order bias parameters $\{b_{\delta^3}, b_{\delta \mathcal{G}_2}, b_{\mathcal{G}_3}, b_{\Gamma_3}\}$ are marginalized over in the likelihood, with Gaussian priors of width $\Sigma_b=1$; (2) the same bias expansion and marginalization but with a much broader prior, $\Sigma_b=1000$; and (3) the second-order bias expansion without any marginalization. Note that case (3) is equivalent to fixing all third-order biases to zero, i.e.\ the limit $\Sigma_b\to 0$. 
In all cases, we fit the 4LPT mocks generated with a cutoff $\Lambda_*=0.1\,h/\mathrm{Mpc}$.

The left panel of \reffig{marg_effect} shows that the setting third-order bias terms to zero yields severely biased results. 
The inferred $b_{\delta^2}$ and $b_{\mathcal{G}_2}$ lie systematically above the RG prediction. Cases (1) and (2) give nearly identical results for $b_\delta$ and $b_{\delta^2}$, indicating that the third-order bias coefficients are likelihood-, not prior-dominated.
The right panel shows the running of third-order bias parameters both considering the full RG (solid lines) and at $\mathcal{O}(\Delta\sigma^2)$ (dashed lines). We see that in that case, higher-order contributions in $\Delta\sigma^2$ are relevant, due to the coupling matrix $s_{ab}^{\rm 1L}$ values. 
By construction, all third-order bias parameters vanish at $\Lambda_*$, but the RG flow generates them at lower cutoffs, with $b_{\delta^3}$ reaching $\mathcal{O}(1)$ values. At this level, the third-order parameters start to affect the estimated running of the first- and second-order parameters.

\section{Validation of $n$-pt measurement with halo catalogs}\label{app:halo}

\begin{figure}
    \centering
    \includegraphics[width=0.45\linewidth]{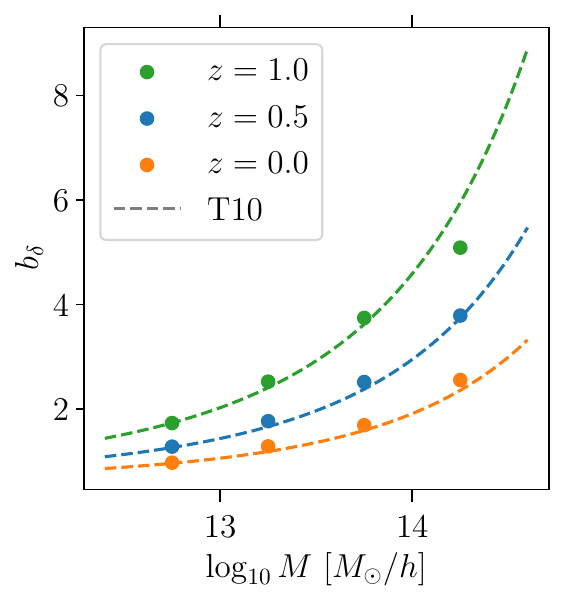}
    \includegraphics[width=0.45\linewidth]{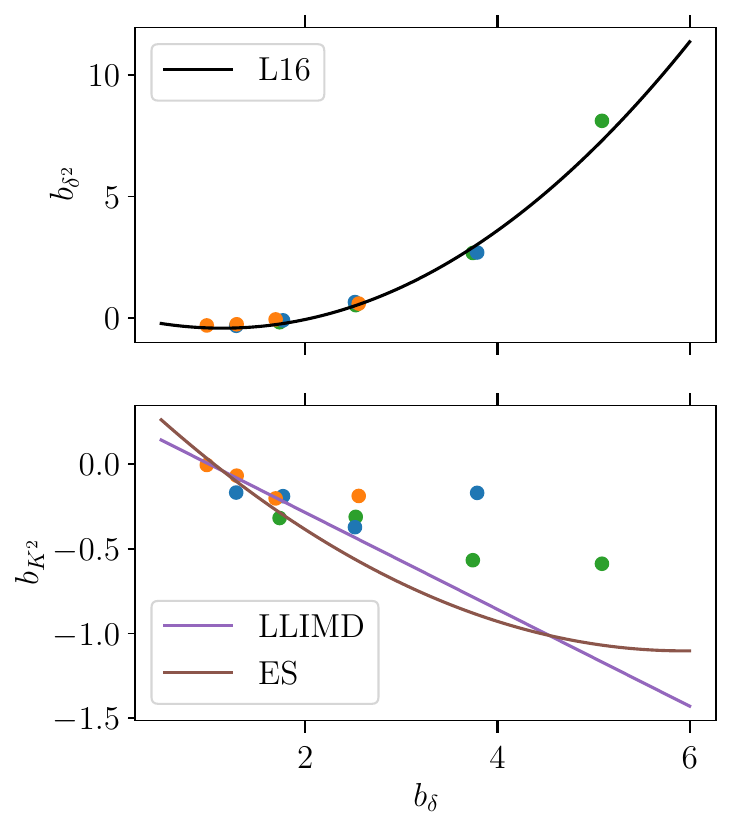}
    \caption{Bias parameters measured with the $n$-point function method on halo catalogs. {Left:} linear bias $b_\delta$ as a function of halo mass at $z=0$ (orange), $0.5$ (blue) and $1$ (green). The dashed lines are the fitting function from \cite{Tinker_2010}, labeled as T10. {Right:} $b_{\delta^2}$ (top) and $b_{K^2}$ (bottom) as functions of $b_\delta$. The black curves in the upper panel show the fitting functions from  \cite{Lazeyras:2015lgp}, labeled as L16. The bottom panel shows the Lagrangian local-in-matter-density (LLIMD) relation (purple) and the simple quadratic fit relation given by the excursion set (ES) formalism (brown) \cite{lazeyras_assembly_2021}.}
    \label{fig:halo_bias_relation}
\end{figure}

As a test of the $n$-point function methodology of \refsec{npt}, we compare the resulting measured linear and quadratic bias parameters with known bias relations, following Fig.~2 of \cite{lazeyras_assembly_2021}. For comparison with \cite{lazeyras_assembly_2021}, we momentarily convert to the $\{b_\delta,b_{\delta^2}, b_{K^2}\}$ basis (see \cite{desjacques_large-scale_2018} for conversion and basis definition), instead of $\{b_\delta,b_{\delta^2}, b_{\mathcal{G}_2}\}$, such that
\begin{equation}
    b_{K^2} = b_{\mathcal{G}_2}\,, \qquad b_{\delta^2}\big|_{K^2} = b_{\delta^2}\big|_{\mathcal{G}_2} - \frac{2}{3}\,b_{\mathcal{G}_2}\,.
\end{equation}
The left panel of \reffig{halo_bias_relation} shows the measured linear bias for different halo masses, compared to the fitting function from \cite{Tinker_2010}. For $b_{\delta^2}(b_\delta)$ we compare in the top right panel of \reffig{halo_bias_relation} our measurements with the fit of \cite{Lazeyras:2015lgp}, that used separate-universe measurements. For $b_{K^2}$, in the bottom right panel, we show predictions for the Lagrangian local-in-initial-density model (LLIDM) $b_{K^2} = -\tfrac{2}{7}(b_\delta - 1)$ relation and the excursion-set (ES) prediction \cite{Sheth_2013}. The behaviours of all the three bias parameters match the results of \cite{lazeyras_assembly_2021} within the expected uncertainties.

\section{Transfer function approach and the missing cutoff}\label{app:transfer_function}
Here we discuss the analytical relation between the transfer function $\beta_O(\vk)$ and the cutoff-dependent bias parameter $b_O^\L$ described by the RG.
We first recap the transfer function approach:
\begin{enumerate}
    \item Shift the operator set, obtained from the matching between the shifted operator kernels \refeq{shifted_operator_kernel} and conventional galaxy bias kernels \refeq{operator_kernel}, with Zel'dovich displacement. The operator set \cite{Schmittfull:2018yuk,ivanov2024fullshapeanalysissimulationbasedpriors} generally considered is \be\label{eq:shifted_operator_set}
    \{\delta_\L(\vq),\delta^2_\L(\vq),\mathcal{G}_{2,\L}(\vq), \delta^3_\L(\vq), \Gamma_{3,\L}(\vq), \mathcal{S}_{3,\L}(\vq)\}
    \ee 
    where $\Gamma_3=\frac{4}{7}\delta\mathcal{G}_2-\frac{4}{7}\frac{\partial_i\partial_j\delta}{\lapl}\frac{\partial_i\partial_j\mathcal{G}_2}{\lapl}$ is the expression derived under the third perturbative order in $\dlin$ instead of the general Galilean operator definition, see App. C in \cite{desjacques_large-scale_2018}, and $\mathcal{S}_3=\bm{s}_2(\vq)\cdot\nabla\delta_1(\vq)$. Notice that all operators are cutoff-dependent due to the lattice filter on $\dlinL(\vq)$.
    
    \item Orthogonalize shifted operators $\tO{O}\to\tObot{O}$ with the Gram-Schmidt rotation matrix, see App. C in \cite{Schmittfull:2018yuk}.
    
    \item Promote the coefficients in front of the orthogonalized shifted operators to free functions of $k$ and project the contributions missing in the model on the existing operators with the optimal transfer function conditions \cite{Schmittfull:2018yuk},
    \be\label{eq:transfer_function}
    \beta_i(k,\L)\equiv\frac{\langle\hdg(\vk) \tObot{O}_{i,\L}(-\vk)\rangle}{\langle \tObot{O}_{i,\L}(\vk)\tObot{O}_{i,\L}(-\vk) \rangle},
    \ee
    where the fact that the basis of operators $\{\tObot{O}\}$ is orthogonal is used. This reduces the model misspecification error and takes account as much nonlinearity as possible. 
    The final transfer function (TF) galaxy density field is 
    \be\label{eq:delta_g_tf}
    \delta_{g,\mathrm{TF}}(\vk)=\sum_{\tO{O}^\bot}\beta_{O}(k,\L) \tO{O}^\bot_\L(\vk).
    \ee 
\end{enumerate}

We argue that the matching relations between the transfer functions $\beta_O(k)$ and the conventional $n$-point function bias coefficients $b_O^{\npt}$ derived perturbatively in \cite{Schmittfull:2018yuk,Obuljen_2023,ivanov2024fullshapeanalysissimulationbasedpriors} are not general identities. Following the concept on unbiased estimation studied in \cite{schmidt_unbiased_2020}, we use the cross-correlations between \refeq{delta_g_tf} with the cutoff-dependent deterministic galaxy bias expansion
\be\label{eq:dgdet_expansion_cutoff}
\dgdet(\vk)=\sum_O b_O^\L O^\mathrm{E}_\L(\vk),
\ee 
to study under what conditions the transfer functions yield physical bias parameters. We assume there is no extra numerical filter in the forward-modeling process except the initial condition.
Notice that \refeq{delta_g_tf} and \refeq{dgdet_expansion_cutoff} are not equal given the difference in construction of the operators involved.

We substitute \refeq{delta_g_tf} and \refeq{dgdet_expansion_cutoff} in \refeq{transfer_function}, and obtain
\be
\beta_i(k,\L)\equiv\sum_{O^{\rm E}_a} \frac{\langle O^\mathrm{E}_{a,\L}(\vk) \tObot{O}_{i,\L}(-\vk)\rangle}{\langle \tObot{O}_{i,\L}(\vk)\tObot{O}_{i,\L}(-\vk)\rangle} b_a^\L.
\ee
The matching coefficient matrix can be defined as
\be\label{eq:matching_cov}
D_{ai}(k,\L)=\frac{\langle O^\mathrm{E}_{a,\L}(\vk) \tObot{O}_{i,\L}(-\vk)\rangle}
{\langle \tObot{O}_{i,\L}(\vk)\tObot{O}_{i,\L}(-\vk)\rangle},
\ee
which relates the finite cutoff $b_O^\L$ to the measured transfer functions. Its $k$-dependence tells us how nonlinear contributions are translated from the bare operators into the fitted transfer functions. One can estimate $D_{ai}^\mathrm{theory}(k,\L)$ perturbatively by substituting the operators $\tObot{O}_\L(\vk)$  and $O^\mathrm{E}_\L$ with their kernel expansion into \refeq{matching_cov}.
This can in turn be compared with measurements to test this prediction.
A systematic derivation of this sort could possibly shed light on the cutoff-dependence of the transfer functions reported in Fig. 12 in \cite{ivanov2024fullshapeanalysissimulationbasedpriors}.

%%%%%%%%%%%%%%%%%%%%%%%%%%%%%%%%%%%%%%%%%%%%
\bibliographystyle{JHEP}
\bibliography{main}

\end{document}